\documentclass[12pt,preprint]{emulateapj}
\begin{document}
\title{Sizes of Lensed Lower-luminosity $z=4$-8 Galaxies from the Hubble Frontier Field Program}
\author{R.J. Bouwens\altaffilmark{1}, G.D. Illingworth\altaffilmark{2}, P.G. van Dokkum\altaffilmark{3}, P.A. Oesch\altaffilmark{4,5}, M. Stefanon\altaffilmark{1}, B. Ribeiro\altaffilmark{1}}
\altaffiltext{1}{Leiden Observatory,
  Leiden University, NL-2300 RA Leiden, Netherlands}
\altaffiltext{2}{UCO/Lick Observatory, University of California, Santa
  Cruz, CA 95064}
\altaffiltext{3}{Department of Astronomy, Yale University, New Haven,
  CT 06520}
\altaffiltext{4}{Department of Astronomy, University of Geneva, Chemin Pegasi 51, 1290 Versoix, Switzerland}
\altaffiltext{5}{Cosmic Dawn Center (DAWN), Niels Bohr Institute, University of Copenhagen, Jagtvej 128, K\o benhavn N, DK-2000, Denmark}
\begin{abstract}
We constrain the rest-$UV$ size-luminosity relation for star-forming
galaxies at $z\sim4$ and $z\sim 6$, 7, and 8 identified behind
clusters from the Hubble Frontier Fields (HFF) program.  The
size-luminosity relation is key to deriving accurate luminosity
functions (LF) for faint galaxies.  Making use of the latest lensing
models and full data set for these clusters, lensing-corrected sizes
and luminosities are derived for 68 $z\sim4$, 184 $z\sim6$, 93
$z\sim7$, and 53 $z\sim8$ galaxies.  We show that size measurements
can be reliably measured up to linear magnifications of
$\sim$30$\times$, where the lensing models are well calibrated.  The
sizes we measure span a $>$1-dex range, from $<$50 pc to $\gtrsim$500
pc.  Uncertainties are based on both the formal fit errors and
systematic differences between the public lensing models.  These
uncertainties range from $\sim$20 pc for the smallest sources to 50 pc
for the largest.  Using a forward-modeling procedure to model the impact
of incompleteness and magnification uncertainties, we characterize the
size-luminosity relation at both $z\sim4$ and $z\sim6$-8.  We find
that the source sizes of star-forming galaxies at $z\sim4$ and
$z\sim6$-8 scale with luminosity $L$ as $L^{0.54\pm0.08}$ and
$L^{0.40\pm0.04}$, respectively, such that lower luminosity
($\gtrsim$$-$18 mag) galaxies are smaller than expected from
extrapolating the size-luminosity relation at high luminosities
($\lesssim$$-$18 mag).  The new evidence for a steeper size-luminosity
relation ($3\sigma$) adds to earlier evidence for small sizes based on
the prevalence of highly magnified galaxies in high shear regions,
theoretical arguments against upturns in the LFs, and other
independent determinations of the size-luminosity relation from the
HFF clusters.
\end{abstract}

\section{Introduction}

The sizes and structures of galaxies contain a significant amount of
information on how they formed.  In particular, the sizes of galaxies
are thought to show a significant proportionality to the size and
structure of the dark matter halos in which these galaxies form.  This
can result in the systematic growth of galaxies with cosmic time
(e.g., Ferguson et al.\ 2004), mirroring the evolution in their dark
matter halos, and to show a strong correlation with galaxy mass (e.g.,
van der Wel et al.\ 2014).  Already, there is significant
observational work, reporting a systematic increase in the sizes of
galaxies with cosmic time (Ferguson et al.\ 2004; Bouwens et
al.\ 2004; Oesch et al.\ 2010; Ono et al.\ 2013; van der Wel et
al.\ 2014; Holwerda et al.\ 2015; Shibuya et al.\ 2015; Suess et
al.\ 2019; Mowla et al.\ 2019a; Whitney et al.\ 2019) and for a
systematic correlation of size with mass (de Jong \& Lacey 2000;
Mosleh et al.\ 2012; Huang et al.\ 2013; van der Wel et al.\ 2014;
Shibuya et al.\ 2015; Mowla et al.\ 2019a,b).  Similarly, significant
work has been done on the structure of galaxies and the structural
evolution, with gas-rich star-forming galaxies showing exponential
profiles and increasingly evolved or dead galaxies showing a de
Vaucouleur profile (e.g., Wuyts et al.\ 2011).  Despite the strong
trends found in the largely parametric analyses listed above of galaxy
sizes and structure, the recovered trends in several other noteworthy
non-parametric analyses (i.e., Curtis-Lake et al.\ 2016; Ribeiro et
al.\ 2017) are considerably less strong, pointing to potential model
dependencies in the interpretation of galaxy structural evolution.

Very low luminosity galaxies lie at one extreme in studies of galaxy
size and structure.  In the local group, most very low luminosity
galaxies ($-10 > M_V > -15$) are found to have half-light radii
ranging from 100 pc to 1 kpc (e.g., McConnachie 2012), with galaxy
size showing a weak correlation with optical luminosity.  At such low
luminosities, the surface brightness of galaxies becomes very low,
ranging from $\sim$24 mag arcsec$^{-2}$ to $\sim26$-27 mag
arcsec$^{-2}$ at $-13$ mag to $-10$ mag, respectively.  Similar
properties are found for dwarf galaxies in the Fornax cluster (Venhola
et al.\ 2017, 2018).  Despite these general trends, when viewed in the
rest-$UV$, dwarf galaxies typically break up into a few distinct
star-forming regions (Overzier et al.\ 2008) which appear as star
cluster complexes.

In this context, it is interesting to study galaxy size and structure
of very low luminosity galaxies in the $z \gtrsim 3$ universe when
galaxies were first forming.  One promising way forward is to make use
of the magnifying effect of gravitational lensing and to combine this
with sensitive, high-resolution views of the distant universe provided
by the Hubble Space Telescope.  Exactly such a view into the distant
universe was made possible with the Hubble Frontier Fields (HFF)
program (Coe et al.\ 2015; Lotz et al.\ 2017).  Sources can be
stretched by large factors along one of their axes.  As we show in
this paper, this stretching can now reliably be measured to linear
magnifications of $\sim$30$\times$, allowing the lensed structure in
systems to be studied at very high spatial resolution (e.g., see
Bouwens et al.\ 2017a for an earlier discussion regarding the
magnification limits from the then-current models).  One significant
earlier example of what can be done was the highly-magnified $z=4.92$
galaxy behind MS1358+62 (Franx et al.\ 1997; Swinbank et al.\ 2009;
Zitrin et al.\ 2011) where star-forming clumps just 200 pc in size
could be partially resolved.

Already, there have been several uses of the HFF observations to
examine the size distribution of extremely faint galaxies.  In an
early study leveraging HFF observations over the first HFF cluster
Abell 2744, Kawamata et al.\ (2015) made use of the data to map out
the distribution of galaxy sizes vs. luminosities, while Laporte et
al.\ (2016) looked further into the sizes of fainter galaxies using
the HFF data over the second and third HFF clusters.  Interestingly,
Kawamata et al.\ (2015) identified a few $\sim-17$ mag
sources\footnote{Specifically HFF1C-i10 and HFF1C-i13 from Kawamata et
  al.\ (2015).} with nominal physical sizes less than 40 pc using
their own lensing model (Ishigaki et al.\ 2015).

In Bouwens et al.\ (2017a), we pursued constraints on the physical
sizes of fainter $>$$-16.5$ mag $z=2$-8 galaxies in the HFF
observations, looking at both (1) the prevalence of sources as a
function of lensing shear and (2) detailed size constraints on sources
in particularly high magnification areas.  These analyses provided
evidence that very low luminosity ($>$$-16.5$ mag) galaxies might have
especially small sizes, i.e., in the range of tens of parsecs to
$\sim$100 pc.  Intriguingly, these sizes are not especially different
from those seen in molecular clouds and star cluster complexes in the
nearby universe (e.g., Kennicutt et al.\ 2003), an idea we develop
further in a companion paper (Bouwens et al.\ 2021: see also Vanzella
et al.\ 2017a, 2019; Renzini 2017; Pozzetti et al.\ 2019).

Kawamata et al.\ (2018) made use of the observations from all six HFF
clusters to measure the size of galaxies behind those clusters, while
calculating the selection efficiency for sources behind the HFF
clusters as a function of size and luminosity.  Combining their size
measurements with the selection efficiencies they computed, they
derived a size-luminosity relationship for lower luminosity galaxies
at $z=6$-9.  They find a steeper size-luminosity relation than what
has been found at higher luminosities (Huang et al.\ 2013; van der Wel
et al.\ 2014; Shibuya et al.\ 2015; Mowla et al.\ 2019a).  42 of the
sources from their $z=6$-9 samples have estimated sizes $\leq$50 pc,
similar to what they had found for some sources in their earlier study
(Kawamata et al.\ 2015) and as had been found by Bouwens et
al.\ (2017a), Vanzella et al.\ (2017a,b; 2019; 2020), and Johnson et
al.\ (2017).

The purpose of the present work is to make use of the observations
from the HFF program to provide an independent measurement of the size
distribution of extremely low luminosity star-forming sources in the
$z=4$ and $z=6$-8 universe, to quantify how the size of these sources
varies with luminosity down to very low luminosities, and finally to
examine the interplay between source sizes and the form of the derived
$UV$ LF.  In doing so, we make use of the clusters from the HFF
program, selecting $z=4$ and $z=6$, 7, and 8 galaxies behind two and
six of them, respectively, and then measure sizes for individual
lensed galaxies.  We discuss the impact of lensing model uncertainties
and incompleteness on the size distributions we derive and discuss the
importance of galaxy sizes on the faint-end shape of the $UV$ LFs.
One caveat in this study is our measurement of source sizes using
rest-$UV$ rather than rest-optical data, and we will discuss this
caveat both in this paper and the companion paper to this study.
Throughout, we assume a standard ``concordance'' cosmology with
$H_0=70$ km s$^{-1}$ Mpc$^{-1}$, $\Omega_{\rm m}=0.3$ and
$\Omega_{\Lambda}=0.7$, which is in good agreement with recent
cosmological constraints (Planck Collaboration et al.\ 2016).
Magnitudes are in the AB system (Oke \& Gunn 1983).

\section{Data Sets and Samples}

In our analysis, we make use of the v1.0 reductions of the HST
observations over all six clusters that make up the Hubble Frontier
Fields program (Coe et al.\ 2015; Lotz et al.\ 2017).  These
reductions include all 140 orbits of HST imaging observations obtained
over each cluster (70 optical/ACS, 70 near-IR/WFC3/IR) plus all
additional archival observations taken over each cluster as a result
of other programs, e.g., CLASH (Postman et al.\ 2012) or GLASS
(Schmidt et al.\ 2014).  All six clusters now have version 3 (v3) and
version 4 (v4) public magnification models are available, including
multiple image systems identified using the full HFF data set and
substantial spectroscopic redshift constraints on multiple image
systems (Mahler et al.\ 2018; Caminha et al.\ 2017; Schmidt et
al.\ 2014; Vanzella et al.\ 2014; Limousin et al.\ 2016; Jauzac et
al.\ 2016; Owers et al.\ 2011).

\begin{figure}
\epsscale{1.15}
\plotone{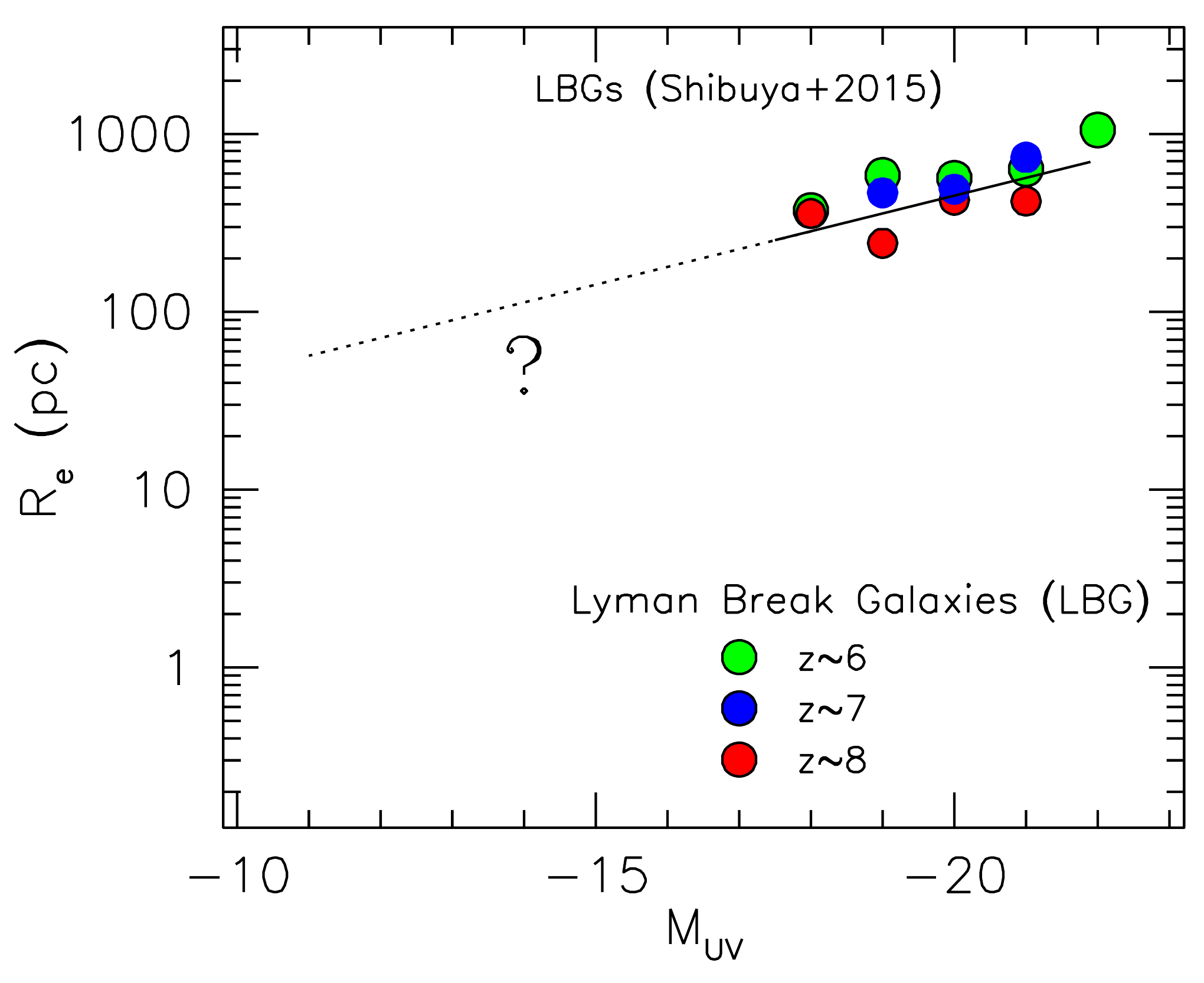}
\caption{Median size vs. luminosity relation of galaxies identified in
  blank field studies, i.e., the XDF/HUDF and CANDELS.  The canonical
  size-luminosity relation is presented using both the Shibuya et
  al.\ (2015) fit results (\textit{black line}) and median sizes at
  $z\sim6$ (\textit{green circles}), $z\sim7$ (\textit{blue circles}),
  and $z\sim8$ (\textit{red circles}).  The black dotted line shows an
  extrapolation of the best-fit Shibuya et al.\ (2015) trend to lower
  luminosities.
\label{fig:msre00}}
\end{figure}

Before constructing catalogs of sources behind these clusters, it is
helpful to attempt to remove both the intra-cluster light and light
from the brightest galaxies which cover considerable ($\sim$40-80\%)
surface area near the center of clusters and make it more difficult to
identify $z\sim6$-8 sources.  The modeling of light from the brightest
foreground galaxies was performed using \textsc{galfit} (Peng et
al.\ 2002), while the intracluster light was modeled using a
median-smoothing algorithm.  While this will be described in detail in
R.J. Bouwens et al.\ (2021, in prep), the algorithm is similar to that
employed by SExtractor (Bertin \& Arnouts 1996) to compute the
background image, i.e., to break the image into $~2''$$\times$$2''$
cells and to compute the median in each.  Then, a median is taken of
the medians in a cell and its 8 nearest neighbors to define the
background level at the center of each cell.  Finally, a spline
interpolation of the median background level at the center of each
cell is performed to estimate the background level across the image.
In Appendix A of Bouwens et al.\ (2017b), this procedure is compared
with similar procedures developed by Merlin et al.\ (2016) and
Livermore et al.\ (2017: see also Shipley et al.\ 2018), finding that
all of these approaches perform similarly well.

After modeling and subtracting light from the foreground cluster and
galaxies from the images, we move onto the selection of faint
high-redshift sources.  Here we restrict our focus primarily to the
selection of sources at $z\sim6$, $z\sim7$, and $z\sim8$ because of
the large number of sources in those samples compared to even higher
redshift selections.  We also make use of results from a faint
selection of $z\sim4$ galaxies to take advantage of the smaller impact
of incompleteness on those selections than at $z\sim6$-8.
Contamination can however be a greater concern for deriving the
size-luminosity relation at $z\sim4$ since evolved galaxies that make
up $z\sim0.3$-0.6 clusters have similar colors to $z\sim5$ galaxies
due to their redshifted Balmer/4000\AA$\,$breaks.  The selection of
$z\sim4$ galaxies can nevertheless be reliably performed if we
restrict ourselves to those behind the two highest redshift clusters
from the HFF program MACS0717 and MACS1149 where the SED shape of the
$z\sim4$ galaxies is sufficiently distinct from galaxies in the
clusters.

The selection of sources in our $z\sim6$, $z\sim7$, and $z\sim8$
samples rely on the following two color criteria and optical
non-detection criteria and is similar to our treatment in Bouwens et
al.\ (2015) and Bouwens et al.\ (2017b).  For our $z\sim6$-7 samples,
we use the criteria:
\begin{eqnarray*}
(I_{814}-Y_{105}>0.6)\wedge(Y_{105}-H_{160}<0.45)\wedge\\
(I_{814}-Y_{105}>0.6(Y_{105}-H_{160}))\wedge\\
(Y_{105}-H_{160}<0.52+0.75(J_{125}-H_{160}))\wedge\\
\textrm{SN}(B_{435}<2)\wedge\\
((\chi_{opt} ^2(B_{435},V_{606})<2)\vee(V_{606}-Y_{105}>2.5))\wedge\\
\textrm{[not in}\,z\sim 8\,\,\textrm{samples]}
\end{eqnarray*}
while for our $z\sim8$ sample, we use the criteria:
\begin{eqnarray*}
(Y_{105}-J_{125}>0.45)\wedge\\
(Y_{105}-J_{125}>0.525+0.75(J_{125}-H_{160})\wedge\\
(J_{125}-H_{160}<0.5)\wedge (\chi_{opt,0.35''} ^2 < 4)\wedge \\
(\chi_{opt,Kron} ^2 < 4)\wedge(\chi_{opt,0.2''} ^2 < 4)
\end{eqnarray*}
where $\wedge$, $\vee$, and S/N represents the logical \textbf{AND}
operation, the logical \textbf{OR} operation, and signal to noise,
respectively.  The $\chi_{opt}^2$ quantity shown above is defined
based on the fluxes in bands blueward of the Lyman break as
$\Sigma_{i} \textrm{SGN}(f_{i}) (f_{i}/\sigma_{i})^2$ and where
$f_{i}$ is the flux in band $i$ in a consistent aperture, $\sigma_i$
is the uncertainty in this flux, and SGN($f_{i}$) is equal to 1 if
$f_{i}>0$ and $-1$ if $f_{i}<0$ (see Bouwens et al.\ 2011).  Three
different apertures are considered for the $\chi_{opt}^2$ parameter,
i.e., a 0.35$''$-diameter aperture, a small scalable Kron aperture,
and a small 0.2$''$-diameter aperture.

Our joint $z\sim6$-7 sample is split into separate $z\sim6$ and
$z\sim7$ galaxy samples based on whether the best-fit photometric
redshift $z_{phot}<6.3$ or $z_{phot}>6.3$, using a similar procedure
to that described in Bouwens et al.\ (2017a) or Bouwens et
al.\ (2017b).  Due to the small size of the Lyman break for $z\sim6$
galaxies, we require that $z\sim6$ sources cannot have an integrated
likelihood of being at $z<4.3$ greater than 35\%.

For our $z\sim4$ sample, we use the following two-color selection
criteria:
\begin{eqnarray*}
(B_{435}-V_{606}>1)\wedge(I_{814}-J_{125}<1)\wedge\\
(B_{435}-V_{606}>1.6(i_{775}-J_{125})+1)\wedge \\
(V_{606}-I_{814}<0.5)
\end{eqnarray*}
Similar to the two color criteria used in Bouwens et al.\ (2021) for
selections of $z\sim4$ galaxies in the HFF parallel fields, we require
that the $V_{606}-I_{814}$ color be bluer than 0.5 mag to avoid
contamination from cluster galaxies.  Sources in our $z\sim4$ sample
are only included brightward of an $H_{160,AB}$ magnitude of 27.3 to
ensure contamination from cluster galaxies is kept to a minimum.  This
is chosen to ensure sufficient S/N in the flux measurements to allow
for a largely clean selection of $z\sim4$ galaxies.  We motivate this
limit in detail in R. Bouwens et al.\ (2021, in prep).

For both our $z\sim6$-8 and $z\sim4$ selections, all bright
($H_{160,AB}<27$) sources with SExtractor stellarity parameters in
excess of 0.9 (where 0 and 1 correspond to extended and point sources,
respectively) are removed.  We also remove sources in cases where the
stellarity parameter is in excess of 0.6 and the HST photometry are
much better with SEDs of low-mass stars ($\Delta \chi^2 > 2$) from the
SpeX library (Burgasser et al.\ 2004) than with a linear combination
of galaxy templates from EAZY (Brammer et al.\ 2008).  Additionally,
we have aimed to be conservative with our selection in that any
sources that lie in particularly noisy regions of the images (e.g.,
where a bright foreground source is subtracted) or overlap with
diffraction spikes are excluded due to the challenges in ascertaining
their reality or accurately characterizing their sizes.

Our $z\sim6$, $z\sim7$, and $z\sim8$ samples from six HFF clusters
contain 184, 93, and 53 sources, respectively, for a total of 330
sources.  Meanwhile, our $z\sim4$ sample identified behind the
MACS0717 and MACS1149 clusters contain 68 galaxies.  In \S4.3, we present
the coordinates and other characteristics of sources in these samples.

\section{Size-Luminosity Relation for Star Forming Galaxies at $z$$\sim$6-8 from Blank Field Studies}

To provide context for the measurements we obtain of the size and
luminosities of faint $z=6$-8 galaxies in the HFF observations (\S4),
it is useful for us to frame the constraints we obtain here for lensed
sources in our fields relative to the sizes of galaxies identified in
an extensive set of blank field studies (e.g., Ferguson et al.\ 2004;
Bouwens et al.\ 2004; Oesch et al.\ 2010; Grazian et al.\ 2012; Huang
et al.\ 2013; Ono et al.\ 2013; Shibuya et al.\ 2015; Holwerda et
al.\ 2015).

The most recent and comprehensive of these determinations is by
Shibuya et al.\ (2015), who conduct size measurements on $\sim$190,000
$z=0$-10 galaxies identified over the XDF/HUDF, the HUDF parallel
fields, the 5 CANDELS fields, and two of the HFF parallel fields.  The
median half-light radius of sources that Shibuya et al.\ (2015)
measure for their $z\sim6$, $z\sim7$, and $z\sim8$ samples is
presented in Figure~\ref{fig:msre00} with green, blue, and red
circles, respectively, and we find that it is well represented by the
following relationship:
\begin{equation}
\log_{10} (r_e/\textrm{pc}) = -0.1(M_{UV}+21) + 2.74
\end{equation}
where $r_e$ is the half-light rest-UV radius in pc and $M_{UV}$ is the
$UV$ luminosity at $\sim$1600\,\AA.  The above size-luminosity relation
is schematic in form, with the intention to present the median
relation for $z\sim6$, 7, and 8.  This relation is included in
Figure~\ref{fig:msre00} as a solid black line over the range where
current observations provide a direct constraint on the relationship
and extrapolated to lower luminosities assuming the same slope
(\textit{dotted line}).  The slope is consistent with the approximate
median slope of the $z\sim6$, $z\sim7$, and $z\sim8$ size-luminosity
relations presented in Figure 10 of Shibuya et al.\ (2015).

The Shibuya et al.\ (2015) size-luminosity relation is fairly typical
of that seen in other studies (Mosleh et al.\ 2012; Huang et
al.\ 2013; van der Wel et al.\ 2014) for luminous galaxies across a
range of redshifts, from $z\sim2$ to $z\sim6$.  However, it is
valuable to recognize that current blank-field HFF observations only
probe the high end of the luminosity range examined in this study.
The HFF studies provide a unique opportunity to extend the analysis to
far fainter luminosities.

\section{Sizes of $z\geq 4$ HFF sources} 

\subsection{Measurement Procedure\label{sec:measurement}}

In fitting the two-dimensional spatial profile of galaxies behind the
HFF clusters to measure sizes, we must account for the substantial
impact that gravitational lensing from the foreground cluster has on
the spatial profile of galaxies.

The relevant quantities in computing the size of a lensed source is
both the total magnification factor $\mu$ and the source shear.  In
Bouwens et al.\ (2017a), we introduced a quantity that we called the
shear factor $S$ which we defined as follows:
\begin{equation}
S = \left\{
\begin{array}{lr}
\frac{1-\kappa - \gamma}{1-\kappa+\gamma}, &
\text{for } \frac{1-\kappa - \gamma}{1-\kappa+\gamma} \geq 1\\
\frac{1-\kappa + \gamma}{1-\kappa-\gamma}, &
\text{for } \frac{1-\kappa - \gamma}{1-\kappa+\gamma} < 1
\end{array}\right.
\end{equation}
where $\kappa$ is the convergence and $\gamma$ is the shear.  The
shear factor $S$ gives the axis ratio a circular galaxy would have due
to the impact of gravitational lensing.

The source magnification $\mu$ can be computed from the convergence
$\kappa$ and shear $\gamma$:
\begin{displaymath}
\mu = \frac{1}{(1-\kappa)^2 - \gamma^2}
\end{displaymath}

The impact of the gravitational lensing on background galaxies is to
stretch sources by a factor $\mu^{1/2} S^{1/2}$ along the major shear
axis and by a factor $\mu^{1/2} S^{-1/2}$ perpendicular to the major
shear axis.

\begin{figure}
\epsscale{1.17}
\plotone{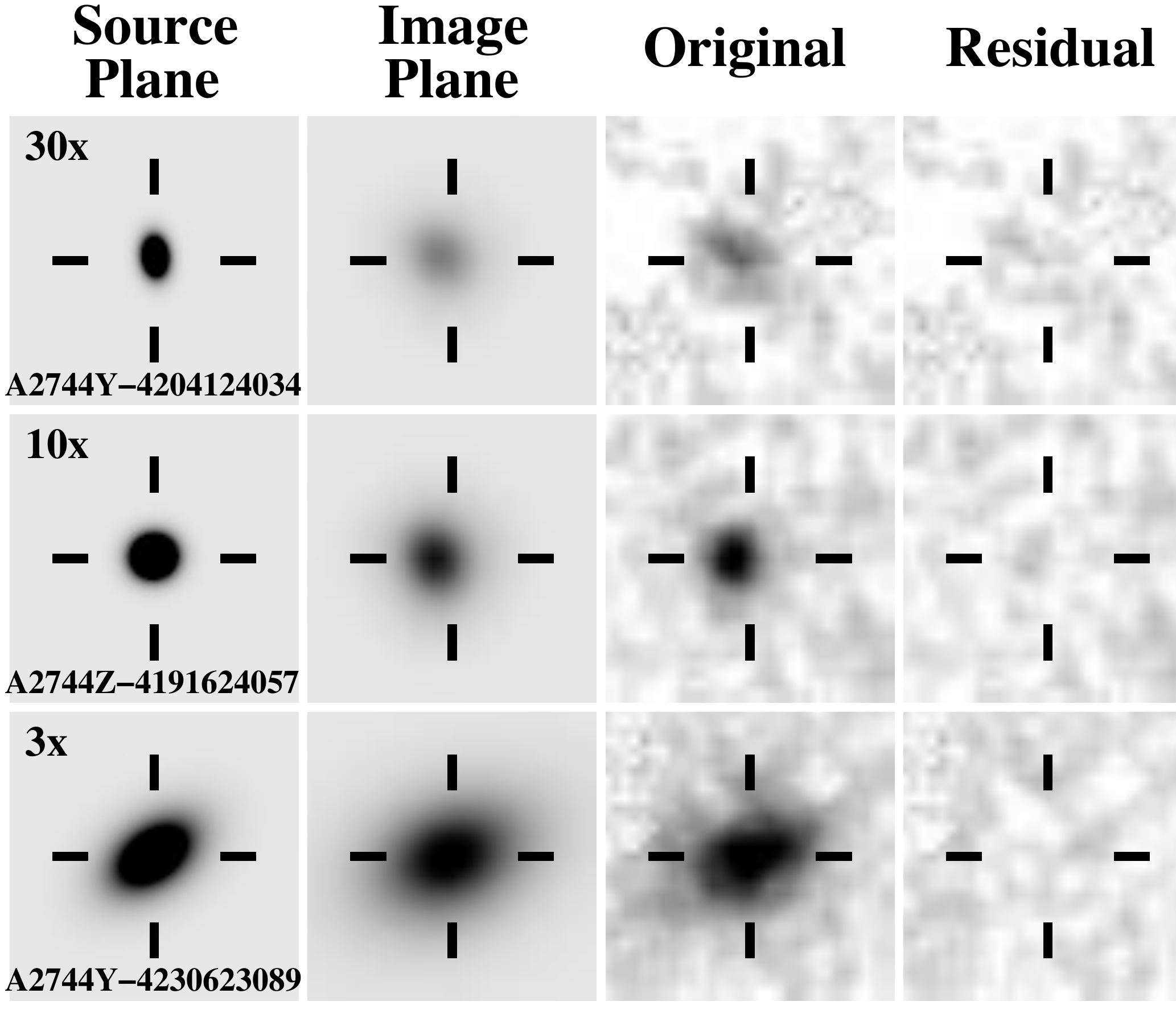}
\caption{Illustration of the typical profile fits used here
  (\S\ref{sec:measurement}) in deriving half-light radii for sources
  lensed by the HFF clusters.  The source-plane models in the leftmost
  column (\textit{shown at the various magnification factors indicated
    in the leftmost postage stamp}) are transformed into the image
  plane and convolved with the PSF to produce the model profiles in
  the image plane (\textit{shown in the second leftmost column}) for
  comparison with the observed two-dimensional images (\textit{second
    rightmost column}) for each source.  The residuals of our profile
  fits are shown in the rightmost column.  Both the observed and model
  images are inverse variance-weighted coadditions of the $Y_{105}$,
  $J_{125}$, $JH_{140}$, and $H_{160}$ images.\label{fig:fit}}
\end{figure}

\begin{figure}
\epsscale{1.17}
\plotone{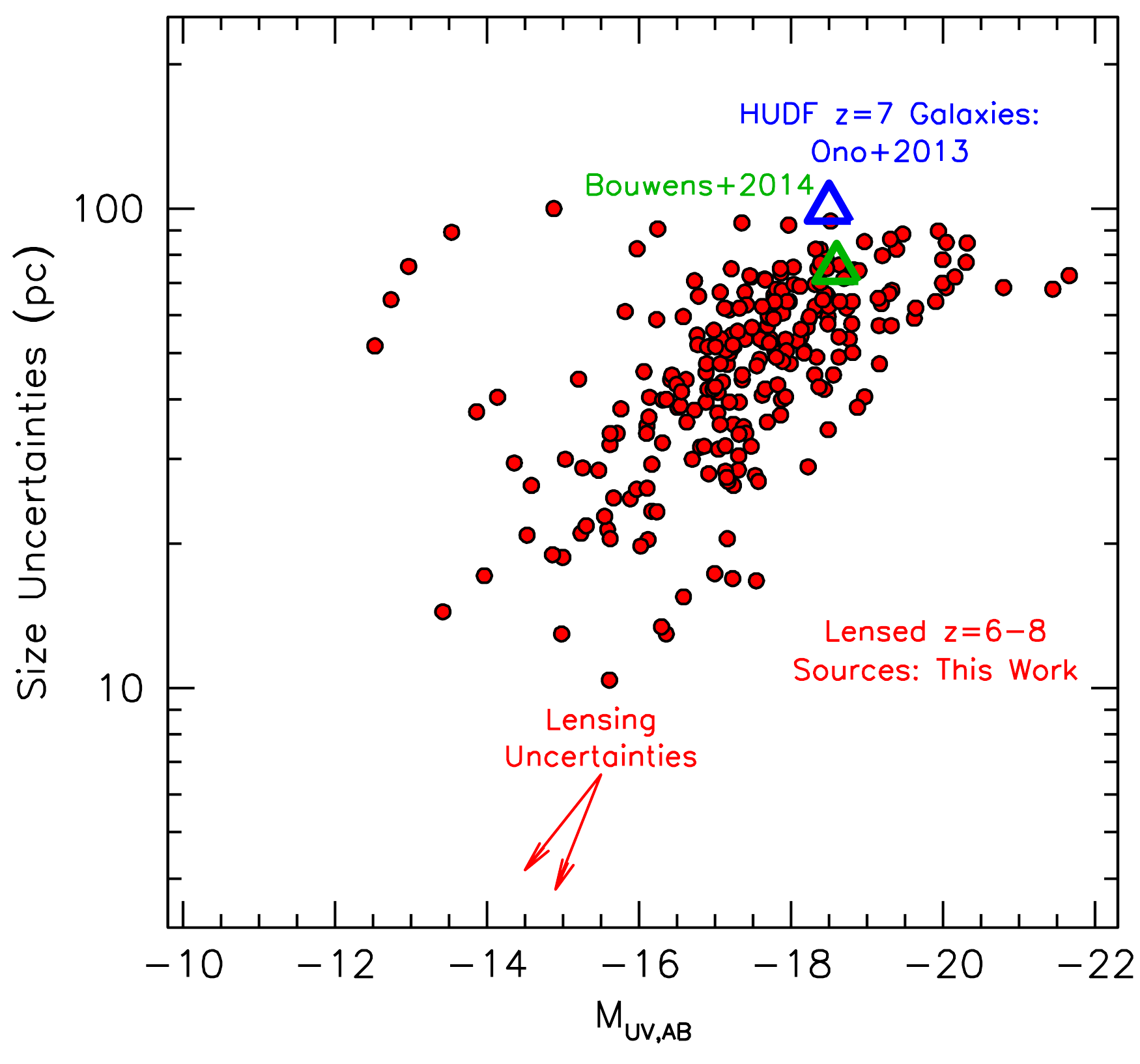}
\caption{Nominal $1\sigma$ accuracy with which source sizes can be
  measured for individually lensed $z=6$-8 sources identified behind
  various HFF clusters vs. the inferred $UV$ luminosity (\textit{red
    circles}).  The accuracy of size measurements is computed by
  adding in quadrature the size uncertainty based on the MCMC fit
  results and the size uncertainty resulting from the unknown lensing
  magnification (based on the dispersion in the lensing models). For
  comparison, we also show the $1\sigma$ uncertainties reported on the
  size measurements of individual $z=7$-8 sources from the HUDF data
  (Ono et al.\ 2013: \textit{open blue triangle}) and for a stack of
  $z\sim7$ sources from the XDF/HUDF (Bouwens et al.\ 2014:
  \textit{open green triangle}).  The red arrows show the range of
  directions sources scatter in $UV$ luminosity and size -- or size
  uncertainties -- due to errors in the magnification models (ranging
  between a $r\propto L$ and $r\propto L^{1/2}$ scaling depending on
  whether magnification is primarily along one or two
  axes).\label{fig:resol}}
\end{figure}

We measure half-light radii of sources via a Markov chain Monte-Carlo
(MCMC) algorithm where we compare the observed two-dimensional profile
with a lensed model profile of a model source with a Sersic radial
profile with its semi-major and semi-minor axes oriented at some
position angle on the sky.  In fitting to the two-dimensional profile,
we coadd the $Y_{105}$, $J_{125}$, $JH_{140}$, and $H_{160}$ images
together after scaling the fluxes in the images to a fixed $f_{\nu}$
frequency and weighting the images by the inverse variance.  No
PSF-matching of the images is performed prior to coaddition.  To
ensure that the fitting is done with a similar composite PSF, the
$Y_{105}$, $J_{125}$, $JH_{140}$, and $H_{160}$ PSFs are similarly
coadded to derive the PSF for the fit procedure.  Thanks to the
similar FWHM of $Y_{105}$, $J_{125}$, $JH_{140}$, and $H_{160}$-band
PSFs (differing by only 5\% in FWHM), differences in the rest-$UV$
colors of star-forming sources at $z\sim6$-8 would only have a small
impact on our size fits.  To illustrate, the effective FWHM for the
PSF of sources with slightly bluer or redder $UV$ colors
($\Delta\beta\sim0.4$) would only differ by 1\%.

We fix the Sersic parameter to 1, motivated by results of Wuyts et
al.\ (2011) who find a predominantly $n=1$ Sersic parameter for
fainter star-forming sources, for self-consistency, and to simplify
the intercomparison of sources within our samples.  For other Sersic
parameters ($n=2$, 3), we would infer larger sizes for sources (by
$\gtrsim$1.5$\times$).  Included amongst the fitting parameters are
the source center, source brightness, source radius, source positional
angle, source axial ratio, and sky background.  The sky background is
refit at each step in the MCMC chain.  This is done by taking the
background which minimizes the square of the residuals.  Lensing is
modeled as magnifying the source by the factor $\mu^{1/2} S^{1/2}$
along the major shear axis and by the factor $\mu^{1/2} S^{-1/2}$
along the minor shear axis.

\begin{figure*}
\epsscale{1.1}
\plotone{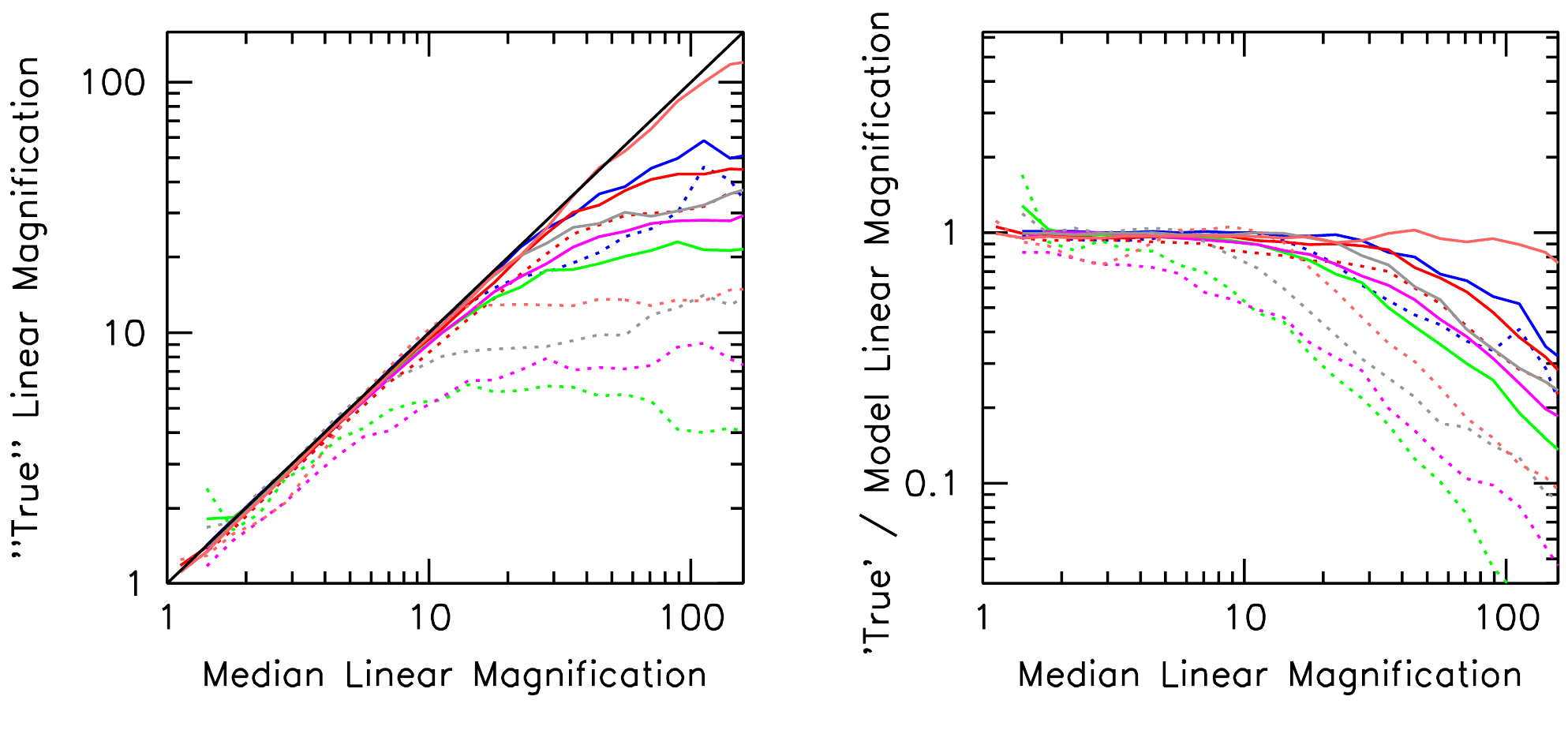}
\caption{An illustration of how well the median linear magnification
  factor likely predicts the actual linear magnification factor.
  (\textit{left}) The plotted solid lines show the linear
  magnification factor of individual parametric models
  (Table~\ref{tab:models}) for Abell 2744 (\textit{red}), MACS0416
  (\textit{blue}), MACS0717 (\textit{green}), MACS1149
  (\textit{magenta}), Abell 370 (\textit{grey}), and Abell S1063
  (\textit{orange}) as a function of the median magnification factors
  from the other parametric models.  In deriving the plotted relation,
  the function is derived for each parametric model individually and
  then a median of the functions is taken.  The dotted lines show the
  relationship, if the median linear magnification factor from the
  parametric models is compared against the median magnification
  factor from the non-parametric models, i.e., \textsc{Grale}, Bradac,
  and Zitrin-LTM (Liesenborgs et al.\ 2006; Sebesta et al.\ 2016;
  Bradac 2009; Zitrin et al.\ 2012, 2015).  The solid black line is
  shown for comparison to indicate the relationship that would be
  present for perfect predictive power for the lensing models.
  (\textit{right}) Similar to the left panel but dividing by the
  quantity plotted along the horizontal axis, i.e., the model linear
  magnification factor.  The linear magnification factors appear to
  have predictive power to factors of $\sim$30 if we assume that the
  parametric lensing models are taken to represent a plausible
  representation of the actual lensing model and $\sim$10 if we assume
  that the non-parametric models are.  This figure is similar in form
  to Figure 3 from Bouwens et al.\ (2017b), though that figure is for
  the total magnification factor.  It is clear that considerable
  caution is required in using results with linear magnification
  factors $\gtrsim$30 ($\gtrsim$10 for non-parametric lensing models).
\label{fig:predict}}
\end{figure*}

Figure~\ref{fig:fit} illustrates our two-dimensional profile fits for
three sources in our catalogs, showing the zoomed model images in the
source plane (\textit{leftmost column}), the PSF-convolved model
images in the image plane after applying the shear (\textit{second
  leftmost column}), the original images (\textit{second rightmost
  column}), and finally the residual image (\textit{rightmost
  column}).

We now describe the magnification factors $\mu$ and shear factors $S$
that we utilize in our analysis.  For the sake of robustness, we do
not rely on the results from a single lensing model, but instead make
use of the median magnification from all available parametric models,
as we and others have done in the past (Bouwens et al.\ 2017b;
Livermore et al.\ 2017).  We have adopted this approach to mitigate
the impact of the individual lensing models losing predictive power in
different ways when the magnification factors from the models become
particularly high, as we illustrated for the linear magnification
factor $\mu^{1/2} S^{1/2}$ in \S\ref{sec:reliable} and previously
demonstrated in Bouwens et al.\ (2017b) for the magnification factor.

The parametric lensing models we utilize for constructing the median
magnification and shear maps include CATS (Jullo \& Kneib 2009;
Richard et al.\ 2014; Jauzac et al.\ 2015a,b; Limousin et al.\ 2016;
Mahler et al.\ 2018; Lagattuta et al.\ 2017), Sharon/Johnson (Johnson
et al.\ 2014), GLAFIC (Oguri 2010; Ishigaki et al.\ 2015; Kawamata et
al.\ 2016), Zitrin-NFW (Zitrin et al.\ 2013, 2015), Keeton (Keeton
2010), and Caminha et al.\ (2016, 2017).  Each of the six HFF clusters
now have highly-refined models available for most, but typically not
all, varieties of model.  Our Abell 2744 median model makes use of 5
of the models (v4.1 of CATS, v4 of Sharon/Johnson, v3 of GLAFIC, v3 of
Zitrin-NFW, v4 of Keeton), our MACS0416 median model makes use of 6 of
the models (v4.1 of CATS, v4 of Sharon/Johnson, v3 of GLAFIC, v3 of
Zitrin-NFW, v4 of Keeton, v4 of Caminha), our Abell S1063 median model
makes use of 5 of the models (v4.1 of CATS, v4 of Sharon/Johnson, v3
of GLAFIC, v4 of Keeton, v4 of Caminha), while our MACS0717, MACS1149,
and Abell 370 median models make use of 4 of the models (v4.1 of CATS,
v4 of Sharon/Johnson, v3 of GLAFIC, v4 of Keeton).
Table~\ref{tab:models} provides a convenient summary of the models we
use.

We also evaluated the performance of the non-parametric models
(\textsc{Grale}: Liesenborgs et al.\ 2006; Sebesta et al.\ 2016,
Brada{\v c}: Brada{\v c} et al.\ 2009; Hoag et al.\ 2017, Zitrin-LTM:
Zitrin et al.\ 2012, 2015; Lam et al.\ 2014; Diego et al.\ 2015a,
2015b, 2016a, 2016b, 2017).  We found that the parametric lensing
models, particularly version 4, perform better in terms of their
predictive power in our tests (see the discussion in \S4.2 and
Figure~\ref{fig:predict}) than the non-parametric models (see also
Meneghetti et al.\ 2017) and we preferentially use the parametric
models for source size determinations.

In computing the magnification and shear factors for the individual
models (to produce the median), we multiply the relevant $\kappa$ and
$\gamma$'s from the aforementioned public models by the ratio of the
distance moduli $D_{ls}/D_{s}$, where $D_{ls}$ is the angular diameter
distance between the lensing cluster and source and the angular
diameter distance to the source, using the best-fit photometric
redshift for the source to compute the distance.  

In this way, we compute the median linear magnification factor
$\mu^{1/2} S^{1/2}$ and $\mu^{1/2} S^{-1/2}$ along the major and minor
shear axes, respectively.  It is worth remarking that these linear
magnification factors appear to be reliable to values as high as 30,
taking the results of \S\ref{sec:reliable} as indicative.  The
direction of the major shear axis is derived using the version 4.1
CATS magnification model, but is fairly similar for the other
parametric lensing models.

Next, we ask how well we can use the HFF lensing clusters to determine
the scale length of faint galaxies to very small sizes.  We can look
to some recent work from HST imaging observations over the eXtreme
Deep Field (XDF) / Hubble Ultra Deep Field (HUDF: Beckwith et
al.\ 2006; Bouwens et al.\ 2011; Ellis et al.\ 2013; Illingworth et
al.\ 2013) to provide some indication.  Ono et al.\ (2013) measure
source sizes for $z\sim7$-8 galaxies at $\sim$$-$19 mag to a $1\sigma$
uncertainty of $\sim$100 pc and at $\sim$$-$18 mag to a $1\sigma$
uncertainty of $\sim$150 pc, corresponding to $\sim$0.1 native pixel.
In Bouwens et al.\ (2014), the sizes of a stack of $z\sim7$ galaxies
are measured to an estimated $1\sigma$ accuracy of 75 pc at
$\sim-$18.5.  If we assume that the median linear magnification
factors are accurate to factors of $\sim$30, this means we can measure
source sizes to 30$\times$ higher spatial resolution over the HFF
clusters as we can over the XDF/HUDF.  This means we can potentially
measure the linear sizes of sources to a $1\sigma$ accuracy of 3-5 pc,
but this is not the case for the typical source.

In Figure~\ref{fig:resol}, we provide the estimated accuracies with
which we can measure sizes for our lensed $z=6$-8 samples vs. $UV$
luminosity.  The accuracy of size measurements is computed by adding
in quadrature the size uncertainty based on the MCMC fit results and
the size uncertainty resulting from the unknown lensing magnification
(based on the dispersion in the lensing models).  This suggests a
typical half-light radius measurement accuracy of 50 pc and $\sim$20
pc for sources at $-18$ mag and $-$15 mag, respectively (as is also
evident from Figure~\ref{fig:resol}).

\subsection{Maximum Linear Magnification Factors to Which the Lensing Models Appear to Be Reliable\label{sec:reliable}}

While magnification models appear to perform quite well in estimating
the true magnification factors behind lensing clusters (Meneghetti et
al.\ 2017) in the median, these models perform the least well in
predicting the magnification factors very close to the critical
curves.  In these high-magnification regions, the total magnification
factors from the models tend to overpredict the actual total
magnification factors quite significantly (e.g., see Figure 3 from
Bouwens et al.\ 2017b), e.g., at $\mu\gtrsim30$.  However, in the new
version 4 models, the predictive power appears to be notably improved
from the results shown in Bouwens et al.\ (2017b) with predictive
power to total magnification factors of $\gtrsim$50 for most of the
HFF clusters.  This represents a significant gain in the utility of
the models.

\begin{deluxetable}{ccc}
\tablecolumns{3}
\tabletypesize{\footnotesize}
\tablecaption{Parametric Lensing Models Utilized (see also \S\ref{sec:measurement})\tablenotemark{a}\label{tab:models}}
\tablehead{\colhead{Cluster} & \colhead{Model} & \colhead{Version}}
\startdata
Abell 2744 & CATS & v4.1 \\
           & Sharon/Johnson & v4 \\
           & Keeton & v4 \\
           & GLAFIC & v4 \\
           & Zitrin/NFW & v3 \\
\\
MACS0416 & CATS & v4.1 \\
         & Sharon/Johnson & v4 \\
         & Keeton & v4 \\
         & GLAFIC & v4 \\
         & Zitrin/NFW & v3 \\
         & Caminha & v4 \\
\\
MACS0717 & CATS & v4.1 \\
         & Sharon/Johnson & v4 \\
         & Keeton & v4 \\
         & GLAFIC & v3 \\
\\
MACS1149 & CATS & v4.1 \\ 
         & Sharon/Johnson & v4 \\ 
         & Keeton & v4 \\ 
         & GLAFIC & v3 \\
\\
Abell 370 & CATS & v4 \\ 
          & Sharon/Johnson & v4 \\ 
          & Keeton & v4 \\ 
          & GLAFIC & v4 \\
\\
Abell S1063 & CATS & v4.1 \\ 
            & Sharon/Johnson & v4 \\ 
            & Keeton & v4 \\ 
            & GLAFIC & v4 \\
            & Caminha & v4 
\enddata
\tablenotetext{a}{See text for a discussion of the models
  used.}
\end{deluxetable}

\begin{deluxetable*}{ccccccc}
\tablecolumns{7}
\tabletypesize{\footnotesize}
\tablecaption{Properties of the Present Compilation $z=6$-8 and $z\sim 4$ Sources over the HFF clusters\tablenotemark{a,b}\label{tab:all}}
\tablehead{\colhead{ID} & \colhead{R.A.} & \colhead{Decl} & \colhead{$M_{UV}$} & \colhead{$\mu$\tablenotemark{c}} & \colhead{$\mu_{1D}$\tablenotemark{d}} & \colhead{$r_e$ (pc)}}
\startdata
A2744I-4242524441  &  00:14:24.25  &  $-$30:24:44.1  & $-$16.7$_{-0.2}^{+1.1}$  & 6.0$_{-1.0}^{+11.3}$  & 6.0$_{-0.8}^{+10.4}$  & 208$_{-141}^{+155}$\\
A2744I-4231724324  &  00:14:23.17  &  $-$30:24:32.4  & $-$17.5$_{-0.3}^{+0.1}$  & 6.3$_{-1.5}^{+0.7}$  & 5.4$_{-1.8}^{+0.9}$  & 143$_{-28}^{+78}$\\
A2744I-4252524255  &  00:14:25.25  &  $-$30:24:25.6  & $-$16.9$_{-0.1}^{+0.0}$  & 2.7$_{-0.1}^{+0.1}$  & 2.5$_{-0.2}^{+0.1}$  & 203$_{-11}^{+20}$\\
A2744I-4226324225  &  00:14:22.63  &  $-$30:24:22.5  & $-$17.7$_{-0.1}^{+0.7}$  & 2.4$_{-0.2}^{+2.1}$  & 2.0$_{-0.2}^{+0.8}$  & 253$_{-84}^{+79}$\\
A2744I-4223024479  &  00:14:22.30  &  $-$30:24:48.0  & $-$17.2$_{-0.2}^{+0.5}$  & 6.7$_{-1.0}^{+3.9}$  & 6.3$_{-0.5}^{+2.3}$  & 122$_{-41}^{+38}$\\
A2744I-4219124454  &  00:14:21.91  &  $-$30:24:45.5  & $-$15.2$_{-0.1}^{+0.3}$  & 7.9$_{-0.9}^{+2.6}$  & 6.9$_{-0.8}^{+2.2}$  & 80$_{-25}^{+25}$\\
A2744I-4197324257  &  00:14:19.73  &  $-$30:24:25.7  & $-$16.3$_{-0.1}^{+0.1}$  & 4.7$_{-0.5}^{+0.5}$  & 3.2$_{-0.6}^{+0.2}$  & 68$_{-5}^{+17}$\\
A2744I-4212524179  &  00:14:21.25  &  $-$30:24:17.9  & $-$16.1$_{-0.0}^{+0.1}$  & 8.2$_{-0.3}^{+0.8}$  & 3.6$_{-0.1}^{+0.4}$  & 124$_{-17}^{+15}$\\
A2744I-4169524527  &  00:14:16.95  &  $-$30:24:52.8  & $-$20.0$_{-0.1}^{+0.0}$  & 1.7$_{-0.2}^{+0.0}$  & 1.5$_{-0.1}^{+0.3}$  & 338$_{-69}^{+56}$\\
A2744I-4169624404  &  00:14:16.96  &  $-$30:24:40.4  & $-$18.0$_{-0.1}^{+0.1}$  & 1.8$_{-0.2}^{+0.2}$  & 1.6$_{-0.0}^{+0.5}$  & 311$_{-99}^{+83}$
\enddata
\tablenotetext{a}{Table~\ref{tab:all} is published in its entirety
  in the electronic edition of the Astrophysical Journal.  A portion
  is shown here for guidance regarding its form and content.}

\tablenotetext{b}{In cases where the median total magnification and
  linear magnification exceeds 50 and 30, resepectively, we quote
  alternative estimates for the absolute magnitude $M_{UV}$ and
  physical size $r_{e}$ with the total and linear magnification fixed
  to 50 and 30, respectively.  This alternate estimates of $M_{UV}$
  and $r_e$ are provided for sources given the challenges in relying
  upon magnification factors in excess of these values
  (\S\ref{sec:reliable}) and elsewhere (Bouwens et al.\ 2017b).}

\tablenotetext{c}{Median magnification factors (and $1\sigma$ uncertainties) derived weighting equally the latest public version 3/4 parametric models from each lensing methodology (\S\ref{sec:measurement}).}
\tablenotetext{d}{$\mu_{1D}$ are the median one-dimensional magnification factors (and $1\sigma$ uncertainties) along the major shear axis $\mu^{1/2} S^{1/2}$ weighting equally the parametric models from each lensing methodology.  This is the same quantity as $\mu_{tang}$ reported by Vanzella et al.\ (2017a).}
\end{deluxetable*}

For the present analysis of sizes, the principal quantity of interest
is not the overall magnification factor, but rather the magnification
along a single spatial dimension.  While the linear magnification
factor was not explicitly considered in the previous analyses of
Meneghetti et al.\ (2017) and Bouwens et al.\ (2017b), it should
broadly correlate with the predictive power of the model magnification
factors.

We can quantify the linear magnification factors to which our size
measurements are reliable in the same way we previously determined the
total magnification factors to which our lensing maps were
sufficiently predictive of the total magnification factors (Bouwens et
al.\ 2017b).  As in that work, we alternatively treat one of the
models as if it represented reality and investigated to what extent
the median linear magnification factors from the other models
reproduced the linear magnification factors from the outstanding
model.

We present the results in Figure~\ref{fig:predict} assuming either
parametric models or non-parametric models provided us with the true
magnification and shear maps.  The results in that figure show that
the gravitational lensing models seem capable of predicting the linear
magnification factors $\mu^{1/2} S^{1/2}$ to values of $\sim$30 and 10
assuming that parametric and non-parametric models, respectively,
represents the truth.  Above these values, the median linear
magnification factor shows a poorer correlation with the linear
magnification factors in individual models.  We note some variation in
predictive power of the lensing models across the six HFF clusters,
with the models for some clusters (e.g., Abell S1063) appearing to be
more reliable than for others (e.g., MACS0717).

If we assume that the parametric lensing models are plausible
representations of the actual lensing model (as the tests of
Meneghetti et al.\ 2017 also suggest), this suggests that we can rely
on the linear magnification factors to values of $\sim$30.  For values
above 30, our results suggests adopting a linear magnification of 30
to be conservative.

\begin{figure*}
\epsscale{1.15}
\plotone{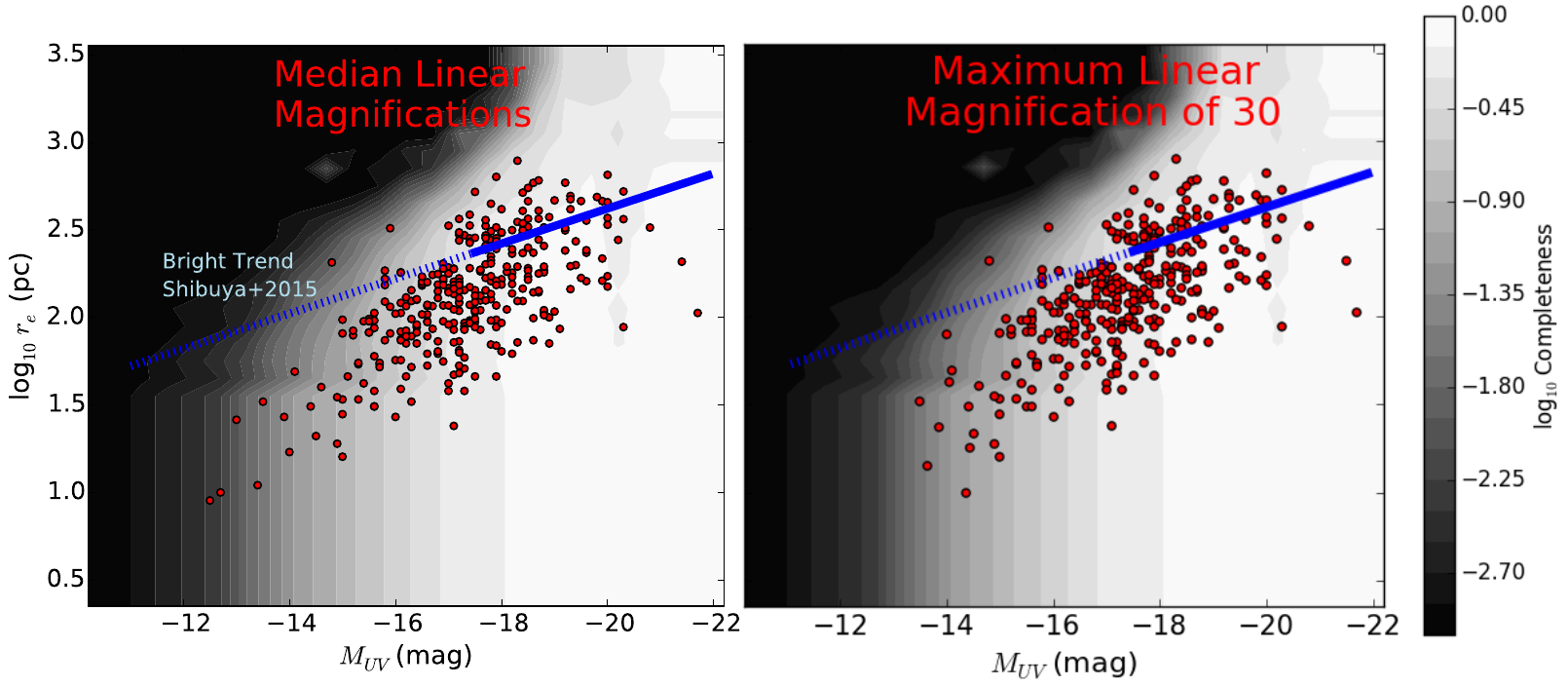}
\caption{Comparison of the distribution of sizes [pc] and luminosities
  for $z=6$-8 galaxies in the HFFs with a canonical size-luminosity
  relationship from blank field studies (\textit{blue solid and dotted
    line in all panels}).  The absolute size in pc is used for the
  left vertical axis here.  The canonical size-luminosity relation for
  galaxies is as in Figure~\ref{fig:msre00}.  For the sizes and
  luminosities of individual lensed $z=6$-8 galaxies, the results
  (\textit{red circles} and $1\sigma$ limits) are based on the median
  magnification factors from the parametric models (\textit{left and
    right panels}).  Also shown, using the grayscale shading in each
  panel, is the relative completeness we estimate for source
  selection, based on extensive source injection and recovery
  experiments.  The relative completeness presented here marginalizes
  over the full area of the HFF clusters and includes the full range
  of magnification factors shown by the clusters
  (\S\label{sec:rawmeasurements}).  Completeness does impact the
  distribution of galaxies in the size-luminosity plane, but the
  overall impact is limited.  For a presentation of the completeness
  using a linear scaling, see Figure~\ref{fig:sizlum_linear} in
  Appendix A.  The right panel is similar to the left panel, but shows
  the size / luminosity distribution after we set those sources with
  total magnification factors $>$50 to 50 and sources with a linear
  magnification factor of $>$30 to 30.  These revised magnifications
  change the $M_{UV}$ and $r_e$ values for the 9 sources where the
  total magnification exceeds 50 and where the linear magnification
  factor exceeds 30.\label{fig:msre0}}
\end{figure*}

\begin{figure}
\epsscale{1.15}
\plotone{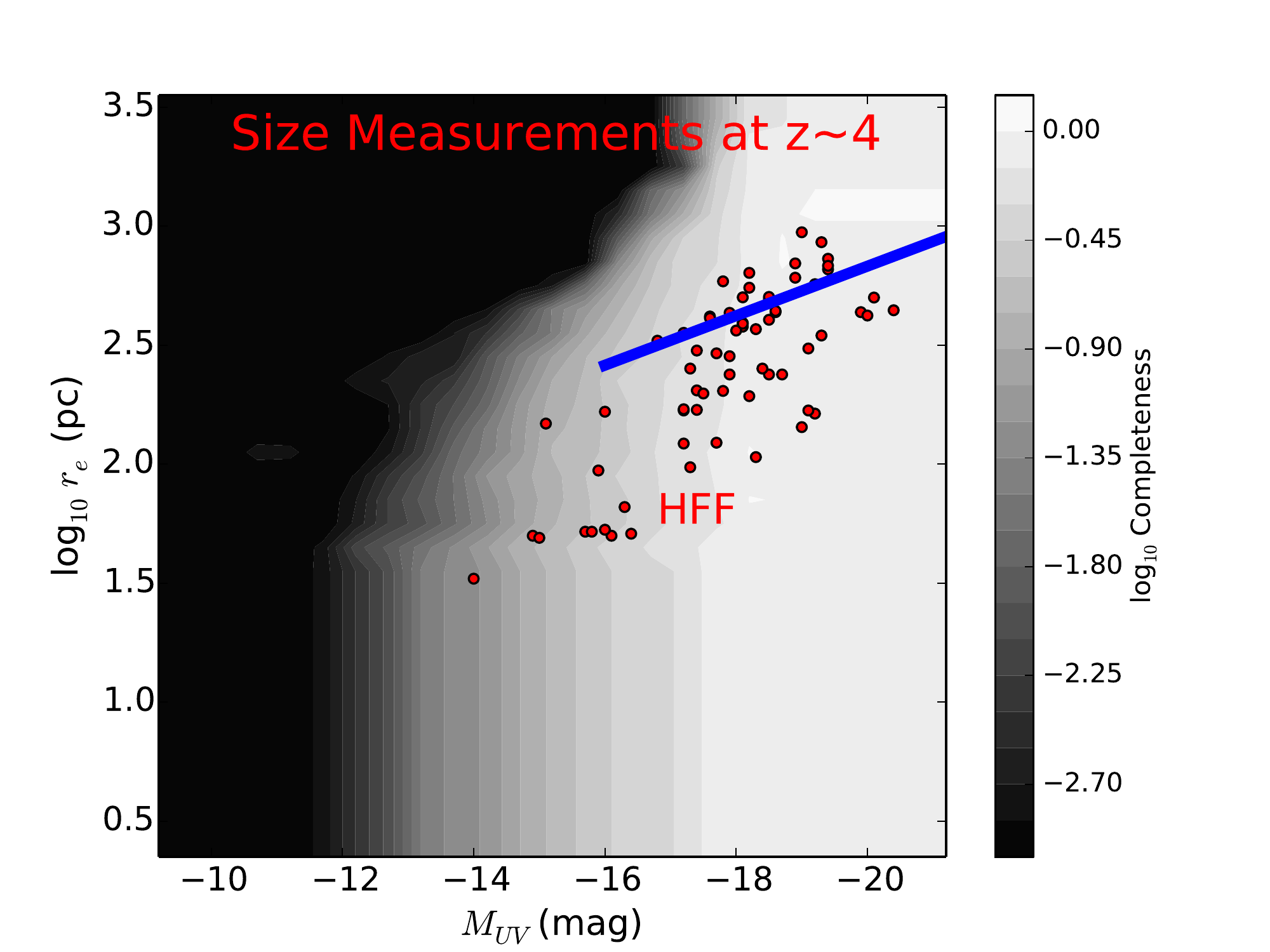}
\caption{Similar to Figure~\ref{fig:msre0} but for sources in our
  $z\sim4$ selection behind the two HFF clusters MACS0717 and
  MACS1149.  For a presentation of the completeness using a linear
  scaling, see Figure~\ref{fig:sizlum4_linear} in Appendix
  A.  \label{fig:msre4}}
\end{figure}

\subsection{Distribution of $UV$ Luminosities and Sizes for $z\geq 4$ Samples\label{sec:rawmeasurements}}

In Table~\ref{tab:all}, we provide the $UV$ luminosities and
half-light radius measurements we derive for the 330 $z=6$-8 sources
and 68 $z=4$ sources utilized in this study, as well as spatial
coordinates, total magnification factors, and linear magnification
factors along the major shear axis.  No sources are excluded as a
result of their S/N falling below some nominal threshold to minimize
the impact of selection effects on the size-luminosity relation we
derive.  The half-light radii we quote are so-called ``circularized''
half-light radii where the radii are equal to the half-light radius
measurement along the semi-major axis multiplied by the square root of
the axial ratios $q$, i.e., $r_e = r_{major,e} q^{1/2}$, where $q =
b/a$, $b$ is the scale along the semi-minor axis, and $a$ is the scale
length along the semi-major axis.

$UV$ luminosities are derived for sources based on their apparent
magnitudes in the $H_{160}$ band and then corrected for the distance
modulus and the total magnification factor.  The distance modulus we
adopt is appropriate for the $z\sim6$, $z\sim7$, and $z\sim8$ samples
we utilize, with median redshift $z\sim5.9$, $z\sim6.8$, and
$z\sim7.9$.  Apparent magnitudes are derived using the same approach
as described in Bouwens et al.\ (2015, 2017b) and rely on scaled Kron
(1980) apertures, which are then corrected to total using point-source
encircled energy distributions appropriate for WFC3/IR (Dressel et
al.\ 2012).  The computed uncertainties on the derived sizes include
both the formal uncertainties on the size fits and the $1\sigma$ error
computed based on the range in linear magnifications predicted by the
parametric lensing models.

In providing estimates for the size and luminosity of lensed HFF
sources in Table~\ref{tab:all}, we rely on very high magnification
factors for some sources.  Fortunately, only a few sources fall in
this regime.  For 7 sources from our 330-galaxy $z=6$-8 sample, the
total magnification estimates exceed 50, and for 5 sources from this
same sample, the linear magnification estimates exceed 30.  For our
$z\sim4$ selection, the total magnification estimates exceed 50 for 4
only sources, and the linear magnification estimates exceed 30 for
only 1 source.  As we have noted, estimates of the magnification
become unreliable at very high values, e.g., when the total
magnification factor exceeds $\sim$50 (Bouwens et al.\ 2017b) and when
the linear magnification factor exceeds $\sim$30 (see \S\ref{sec:fit}
and Figure~\ref{fig:predict}).

Given the lower reliability of magnification factors that exceed these
values, we provide alternative estimates of the sizes and luminosities
for sources in these cases.  For these alternate estimates, we take
the total magnification factor to be 50 for all sources where the
formal estimates exceed 50 (based on a median of the models).
Similarly, we take the linear magnification factor to be 30 for all
sources where the formal estimates exceed 30.  For sources where one
of the two magnification estimates does not exceed these thresholds,
we simply adopt the median magnification, total or linear, from the
models.

To interpret the observed size-luminosity distribution, it is
essential we account for the impact of surface brightness selection
effects on the composition of our sample.  Surface brightness
selection effects are important when the surface brightnesses of
sources lie close to the selection limit of our samples (e.g., Bouwens
et al.\ 2004; Oesch et al.\ 2015; Taghizadeh-Popp et al.\ 2015;
Kawamata et al.\ 2018).  In the extreme case that the selection effect
becomes dominant, it can cause selected sources to show a relatively
fixed range of surface brightnesses ($\propto$$L/r^2$) and for source
size $r$ to depend on radius as $L^{1/2}$ (e.g., Bouwens et
al.\ 2017a; Ma et al.\ 2018).

\begin{figure*}
\epsscale{1.16} \plotone{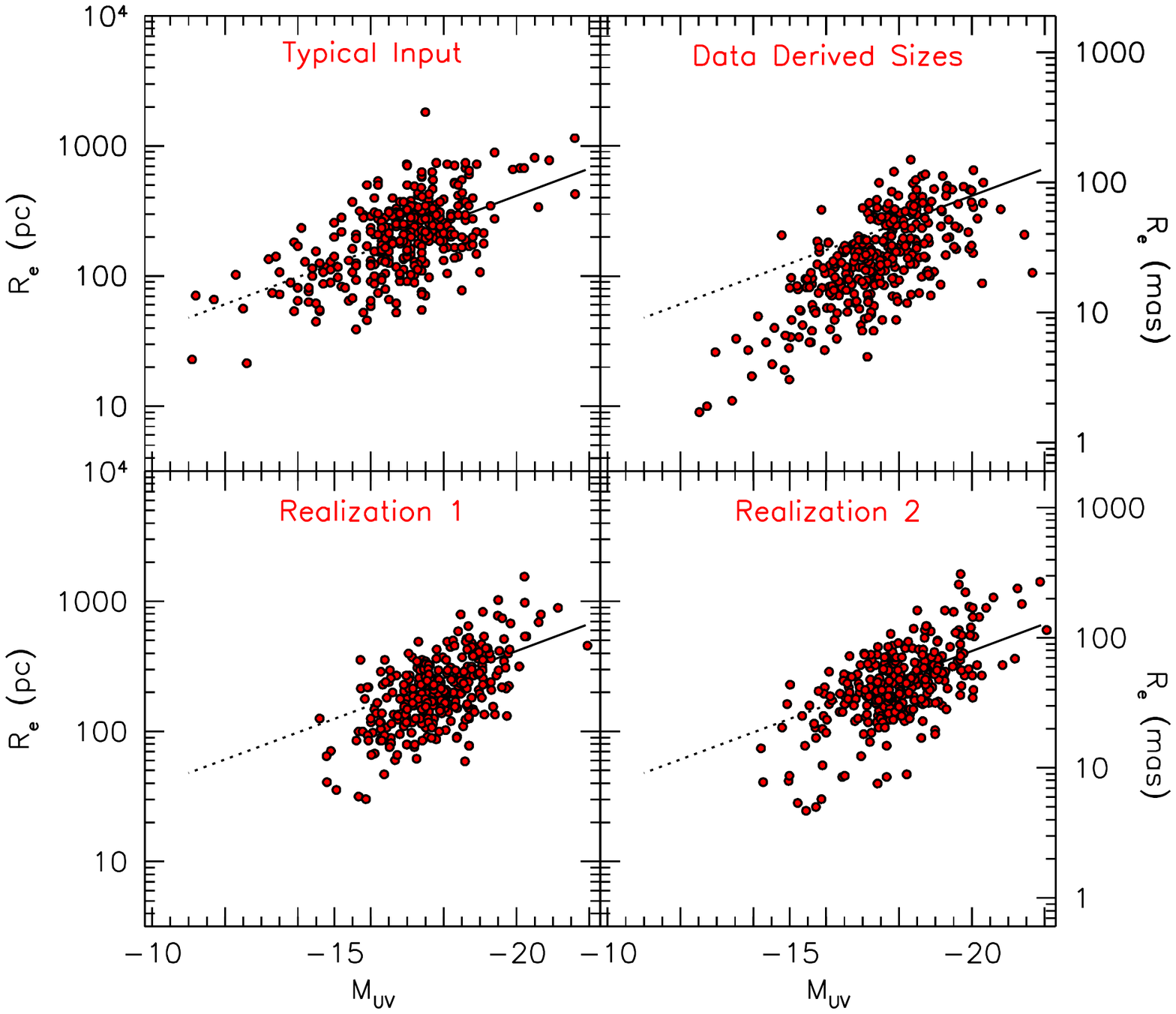}
\caption{Comparison of the observed distribution (\textit{upper
    right}) of lensed $z=6$-8 sources we derive in size and luminosity
  (\textit{red circles}) from our forward-modelling procedure relative
  to the distribution extrapolating the Shibuya et al.\ (2015)
  relation (\textit{solid and dotted lines}) to lower luminosities
  (\textit{upper left}), and two realizations of the size-luminosity
  distribution incorporating the impact of both lensing uncertainties
  and surface brightness selection effects (\textit{lower left and
    right}).  The points are plotted here using their absolute sizes
  in pc; the mas scale on the right vertical axis is indicative of the
  angular sizes at $z\sim7$.  Through comparisons of the lower two
  panels to the upper left panel, it should be clear what a dramatic
  impact both lensing uncertainties and selection effects have on the
  recovered size-luminosity distribution.  While our simulations
  predict that some sources will appear to have very small sizes as a
  result of these effects, we find an even larger number of sources in
  the real observations, suggesting that a fraction of the sources do
  indeed have such small sizes.\label{fig:realizations}}
\end{figure*}

\begin{figure*}
\epsscale{1.09} \plotone{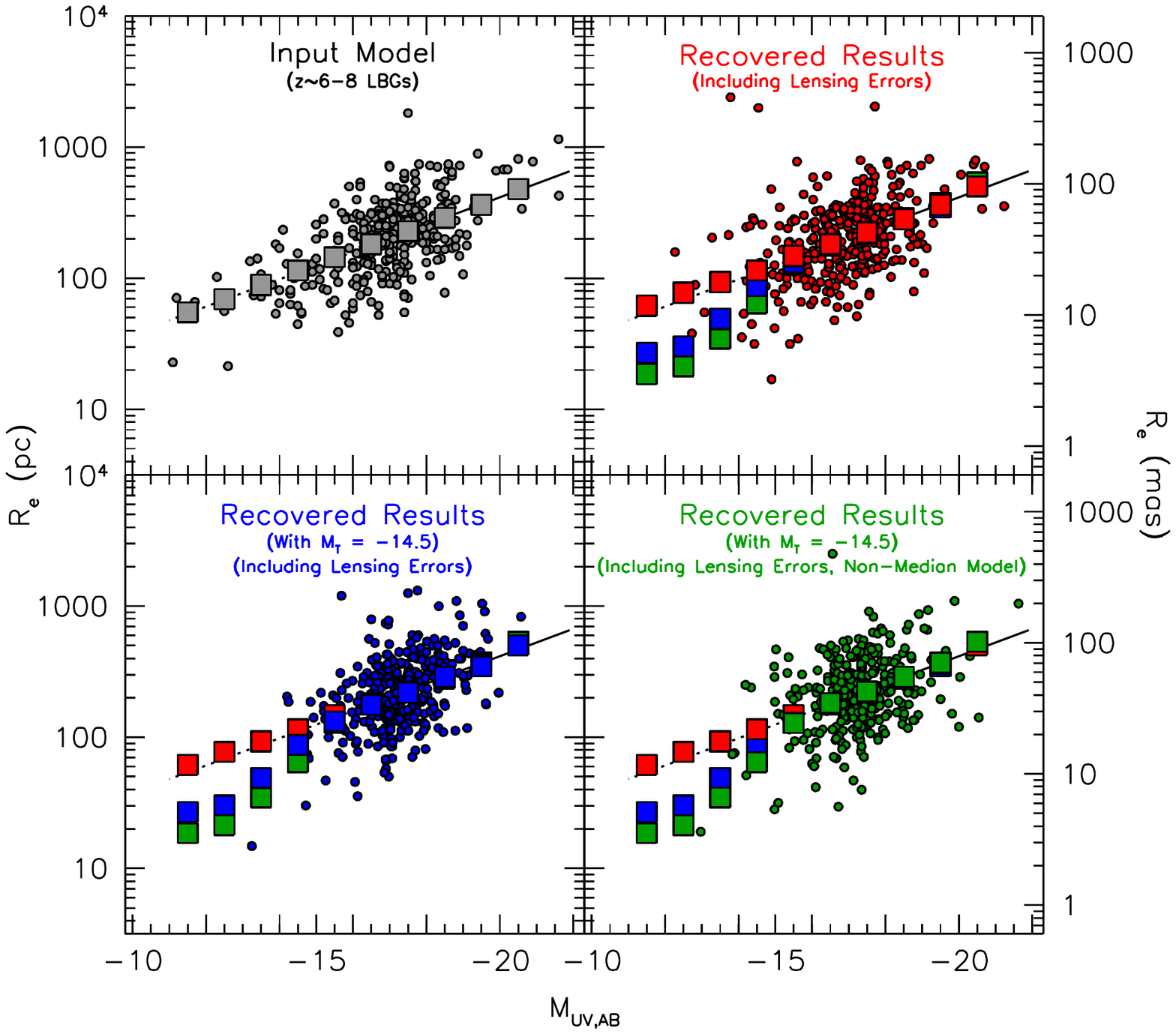}
\caption{Illustration of how errors in the lensing models are expected
  to impact the size-luminosity distribution using Monte-Carlo
  simulations.  The points are plotted here again using their absolute
  sizes in pc; the mas scale on the right vertical axis provides the
  corresponding angular sizes for sources at $z\sim7$.  The upper left
  panel shows an input distribution of sizes and luminosities for a
  $\sim$330 source sample (\textit{shown with the gray solid circles})
  using the size-luminosity relation from Shibuya et al.\ (2015) and
  including the 0.24-dex intrinsic scatter found by Shibuya et
  al.\ (2015).  The upper right and lower left panels show the
  distribution of sizes and luminosities treating the \textsc{CATS}
  models as the truth and recovering the results using the median
  magnification and shear maps from the parametric lensing models.
  The lower right panel shows the distribution treating the
  \textsc{CATS} models as the truth but recovering the results using
  the Keeton magnification and shear maps.  In the upper two panels,
  the input LF adopted for the simulation has a faint-end slope of
  $-2.05$ and no turn-over.  In the lower two panels, the input LF is
  taken to have a faint-end slope of $-2.05$ and turn-over at $-14.5$
  mag.  The large gray, red, blue, and green squares show the median
  size at a given $UV$ luminosity before and after applying the median
  magnification and shear maps in the upper left, upper right, lower
  left, and lower right panels, respectively.  \textit{The median
    sizes are derived based on 100$\times$ more sources than are shown
    on the figure.}  No selection effects are included in these
  idealized results.  The median results shown with the large red,
  green, and blue squares in the upper right, lower left, and lower
  right panels are replicated in the other panels to facilitate
  intercomparisons.  Errors in the lensing models alone can introduce
  significant scatter in the recovered sizes and $UV$ luminosities,
  but do not appreciably impact the recovered median size measurements
  brightward of $-$15 mag and only minorly in the case of the LF with
  no turn-over (\S\ref{sec:lensingmodel}).  Nevertheless, in the case
  of a LF with a turn-over, errors in the magnification model drive
  the results towards artificially small sizes.\label{fig:lenserror}}
\end{figure*}

To compute the completeness distribution, we have run a substantial
set of Monte-Carlo simulations, where we have injected artificial
sources into our simulated images, incorporating the impact of
gravitational lensing from one of the latest HFF public models (using
the \textsc{CATS} deflection maps), and then rerunning our procedures
for source selection.  Intrinsic sizes (from 7.5 mas to 600 mas in
half-light radii) and luminosities (from $-21$ to $-12$ mag) were
considered for the injected sources in the simulations.  Exponential
profiles were assumed for sources in the simulations to match with
that generally found for the spatial profiles of faint star-forming
galaxies in various extragalactic fields (Hathi et al.\ 2009; Wuyts et
al.\ 2012).

In the left panel of Figure~\ref{fig:msre0}, we show the measured
sizes and estimated luminosities of lensed sources in our $z=6$-8
samples in relation to the derived and extrapolated size-luminosity
relation from blank field studies.  We also include on this panel the
estimated completeness we estimate from our source recovery
experiments for sources with a given size and luminosity.  The plotted
completeness, shown in grey scale, marginalizes over the full area of
the HFF clusters and includes the full range of magnification factors
shown by the clusters.  Figure~\ref{fig:msre4} shows the distribution
of our $z\sim4$ galaxy sample in size and luminosity, while showing in
grey scale the estimated completeness.  To provide a separate
illustration of where sources fall in size and luminosity relative to
the estimated completeness (but this time shown using a linear
scaling), we have included Figures~\ref{fig:sizlum_linear} and
\ref{fig:sizlum4_linear} in Appendix A.

Of particular relevance in establishing the form of the
size-luminosity relation at lower luminosities is how incomplete
source selections become, especially adopting a simple extrapolation
to the Shibuya et al.\ (2015) relation.  Galaxies following the
Shibuya et al.\ (2015) size-luminosity relation and having absolute
magnitudes faintward of $-15.5$ mag and especially $-14.5$ mag would
suffer from a significant level of incompleteness in our selections.
It is therefore quite clear that we need to account for selection
completeness in attempting to derive the size luminosity relation.  In
the limiting case that a selection is dominated by surface brightness
selection effects, the recovered source size distribution would tend
to scale as $L^{0.5}$ (e.g., Bouwens et al.\ 2017a; Ma et al.\ 2018).

As many as 7 sources from our $z=6$-8 sample and 4 sources from our
$z\sim4$ sample exceed total magnification factors of 50, where the
magnification models become less reliable.  To gain insight into how
high-magnification sources impact the distributions of galaxies in the
size-luminosity plane, we present in the right panel of
Figure~\ref{fig:msre0} the source sizes and luminosities for our
$z=6$-8 sample, if we set the maximum linear magnification factor and
magnification factor to be 30 and 50, respectively, as we do for the
alternate estimates for the sizes and luminosities in
Table~\ref{tab:all}.  As expected, the distribution shown in the right
panel extends less towards very small sizes and low luminosities.
Reassuringly, the distribution does not differ greatly from what is
shown for the full 330 galaxy sample in the left panel.

Our size measurements shown in the panels of Figure~\ref{fig:msre0}
and \ref{fig:msre4} clearly suggests a steep size-luminosity relation.
The distribution appears to be steeper than what is found for bright
field galaxies represented by the measured solid and dotted
extrapolated blue line, as per Figure~\ref{fig:msre00}.  Reassuringly,
at the bright end ($<$$-$17 mag), lensed sources in our samples
scatter around the expected measured sizes from the blank-field
studies shown in Figure~\ref{fig:msre00}, where the half-light radius
scales with luminosity as $L^{0.26\pm0.03}$.

\subsection{Using Forward Modeling to Characterizing the Size-Luminosity Relation at $z\geq 4$\label{sec:lensingmodel}}

The purpose of this section is to describe our procedure for forward
modeling the apparent distribution of galaxies in the size-luminosity
plane.  Such a modelling is required to incorporate the multiple
effects that can have a significant impact on the recovered
distribution of galaxies in the distant universe for comparisons with
the observed distributions.  Important effects to consider as part of
this modeling include surface brightness selection effects and
uncertainties in the lensing models.

In modeling the size-luminosity distribution, we adopt a $UV$ LF with
a characteristic luminosity $M_{UV} ^{*}$ of $-$21 mag and a faint-end
slope of $-2.05$ and $-1.7$ in modeling our selections of galaxies at
$z\sim6$, 7, and 8 and $z\sim4$, respectively, consistent with the
Schechter parameters derived at these redshifts (e.g., Bouwens et
al.\ 2021).  We furthermore adopt a slope, scatter $\sigma$, and
offset to the size-luminosity relation $r_e = r_0
(L_{UV}/L_0)^{\alpha}$ where $\alpha$ corresponds to the slope of the
size-luminosity relation and $r_0$ corresponds to the offset.  We take
the $UV$ luminosity $L_0$ to be 5.3 $\times$10$^{10}$ $L_{\odot}$
(corresponding to $-$21 mag, assuming a rest-frame wavelength of
1600\AA$\,$for $UV$ luminosity measurements).

Assuming the aforementioned LFs and size-luminosity relations, we
create a Monte-Carlo catalog of sources.  We then apply the selection
functions we computed in \S\ref{sec:rawmeasurements} and as shown in
Figure~\ref{fig:msre00}.  Sources are retained in our Monte-Carlo
catalog if a random number uniformly distributed between 0 and 1
exceeds the completeness of a source at a given absolute magnitude,
physical size, and magnification factor.

We then perturbed the inferred size and luminosity of sources in our
catalog to account for both measurement uncertainties and
uncertainties in the total and linear magnification factors of
sources.  We take the measurement uncertainties to be identical to
that found in the observations for sources of a given apparent
magnitude and size.  To derive perturbed total magnitudes for sources,
we assumed the CATS v4.1 lensing models represent the truth, computed
apparent magnitudes for sources, and then derived absolute magnitudes
for sources using a median of the other parametric lensing models.
The choice of the CATS models for this purpose rather than one of the
other parameteric lensing models (Sharon/Johnson, Keeton, or GLAFIC)
was arbitrarily made.

To derive the perturbed sizes, we first calculated the apparent
angular size of sources based on the linear magnification factors
computed from the CATS v4.1 lensing model and then computed the
physical sizes of sources according to the median linear magnification
factors from the other parametric models.  To ensure that we were
accurately accounting for the impact of the linear magnification
factors on the inferred sizes, we remeasured the sizes of $z\sim4$ +
$z\sim6$-8 galaxies in our catalogs using various linear magnification
factors and determined the impact of the magnification factors on the
derived sizes.

As an illustration of the size-luminosity distribution we derive based
on the above forward-modeling procedure and how it compares to both
the input and observed distribution at $z\sim6$, 7, and 8, we provide
two examples in the lower panels of Figure~\ref{fig:realizations}.
For these two examples, we use an extrapolated version of the
size-luminosity distribution from Shibuya et al.\ (2015) to lower
luminosity and a faint-end slope $\alpha$ of $-2.05$.  Note how
different the forward-modelled size-luminosity distribution is from
the input distribution and thus the significant impact that errors in
the lensing model and incompleteness have on this distribution.

In light of the dramatic differences between the forward-modelled and
input size-luminosity distribution that we see in
Figure~\ref{fig:realizations}, it is interesting to look at the impact
that uncertainties in the lensing magnification factors alone have on
the result.  We illustrate this in Figure~ \ref{fig:lenserror}.  The
upper left panel in this figure shows the input distribution of sizes
and luminosities for a LF model with a faint-end slope of $-2.05$,
while the upper right panel shows the recovered distribution after
using a median model to interpret the mock data set.  The lower left
and right panels show the recovered distribution for a LF model with a
faint-end slope of $-$2.05, but a faint-end turn-over at $-14.5$ mag
following the functional form for a turn-over given in Bouwens et
al.\ (2017b).  The recovery is performed using the median
magnification and shear maps in the lower left panel and using the
Keeton magnification and shear maps in the lower right panel.  The
Keeton magnification maps are utilized for this exercise because they
are available for all six HFF clusters and make use of a different
lens modeling software than the \textsc{CATS} maps.  The large solid
squares show the median half-light radius recovered per 1-mag bin of
$UV$ luminosity. (shown in red, red, blue, and green in the upper
left, upper right lower left, and lower right panels, respectively and
reproduced in the other panels to facilitate intercomparisons).  These
median sizes are derived based on 100$\times$ more sources than shown
on the figure, to more clearly indicate the impact of the lensing
uncertainties.  To focus on the impact of lensing errors alone, no
selection effects are incorporated in the upper right or lower panels.

Comparisons of the upper-left panel from Figure~\ref{fig:lenserror}
with the other panels in the same figure show the impact that
uncertainties in the lensing model have on the recovered sizes and
luminosities for specific sources.  While uncertainties in the lensing
model have an impact on the inferred luminosity and size of individual
sources, these uncertainties only seem to have an impact on the median
inferred sizes faintward of where there is a turn-over in the input
LF.  In the case where no turn-over exists, the most salient impact of
the uncertainties is to increase the apparent width of the size
distribution at a given $UV$ luminosity.  We can thus conclude that
there are two effects that impact the size-luminosity realizations
shown in the two lowest panels of Figure~\ref{fig:realizations}, a
surface brightness selection effect and a broadening in the
size-luminosity distribution owing to uncertainties in the lensing
models.

\begin{figure}
\epsscale{1.15}
\plotone{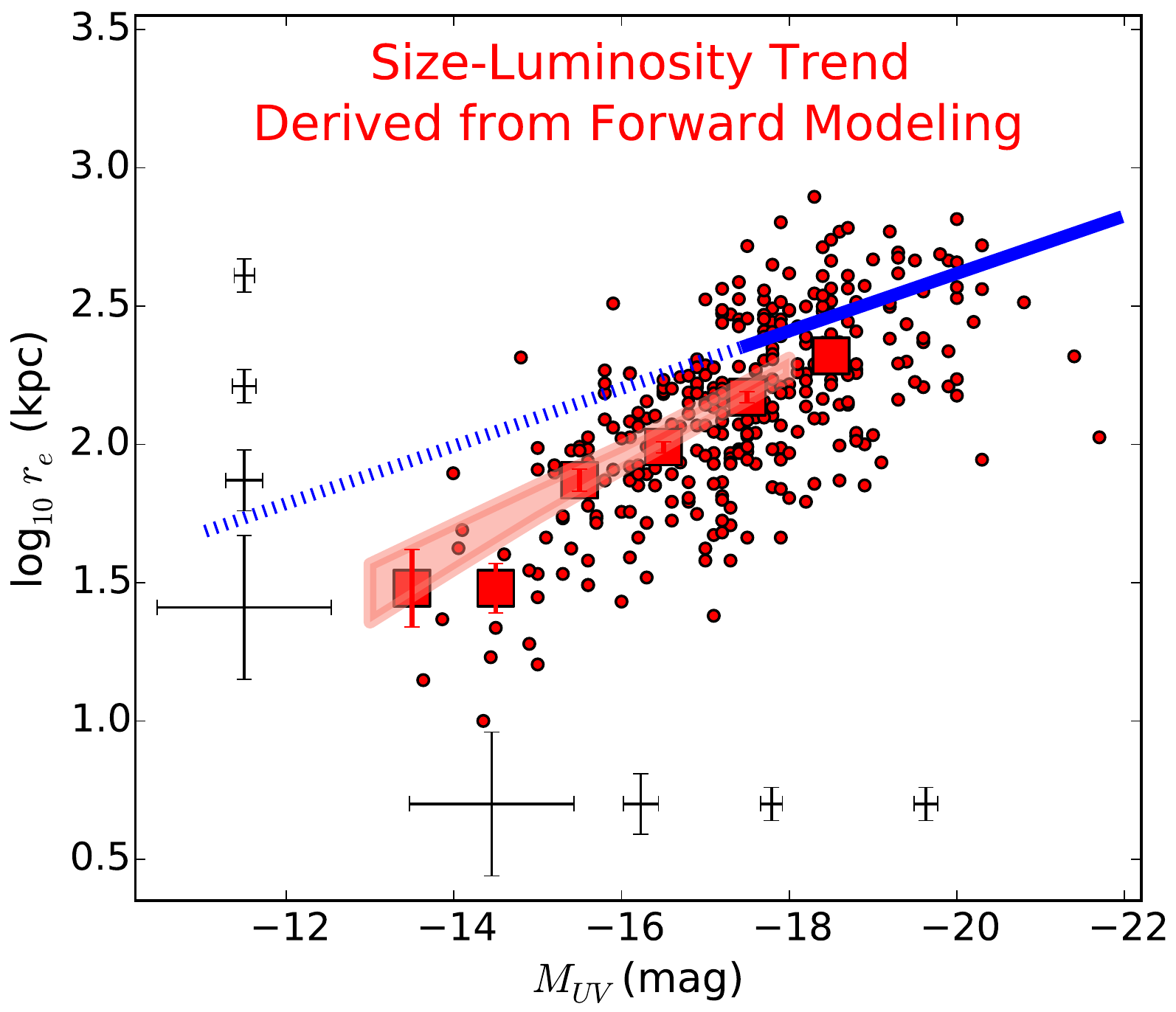}
\caption{Best-fit size-luminosity relation (\textit{shaded red
    region}) derived for galaxies in our $z=6$-8 selection
  (\S\ref{sec:fit}) relative to the Shibuya et al.\ (2015)
  size-luminosity relation relevant for the most luminous $z=6$-8
  galaxies.  The large red squares show the median measured half-light
  radius per 1-mag $UV$ luminosity bin.  The errorbars on the red
  squares are $1\sigma$.  The error bars in the left and lower region
  of this figure show the typical $1\sigma$ errors on the inferred
  sizes and $UV$ luminosities vs. half-light radii and $UV$
  luminosity, respectively.  The size-luminosity relation we infer for
  lower-luminosity $z=6$-8 galaxies is steeper than for higher
  luminosity galaxies.  The steepness of the size-luminosity relation
  derived from our measurements contrasts strikingly with the
  shallower slope of the Shibuya et al.\ (2015) relation at higher
  luminosities, and its extrapolation to lower
  luminosities.\label{fig:sizefit7}}
\end{figure}

\begin{figure}
\epsscale{1.15}
\plotone{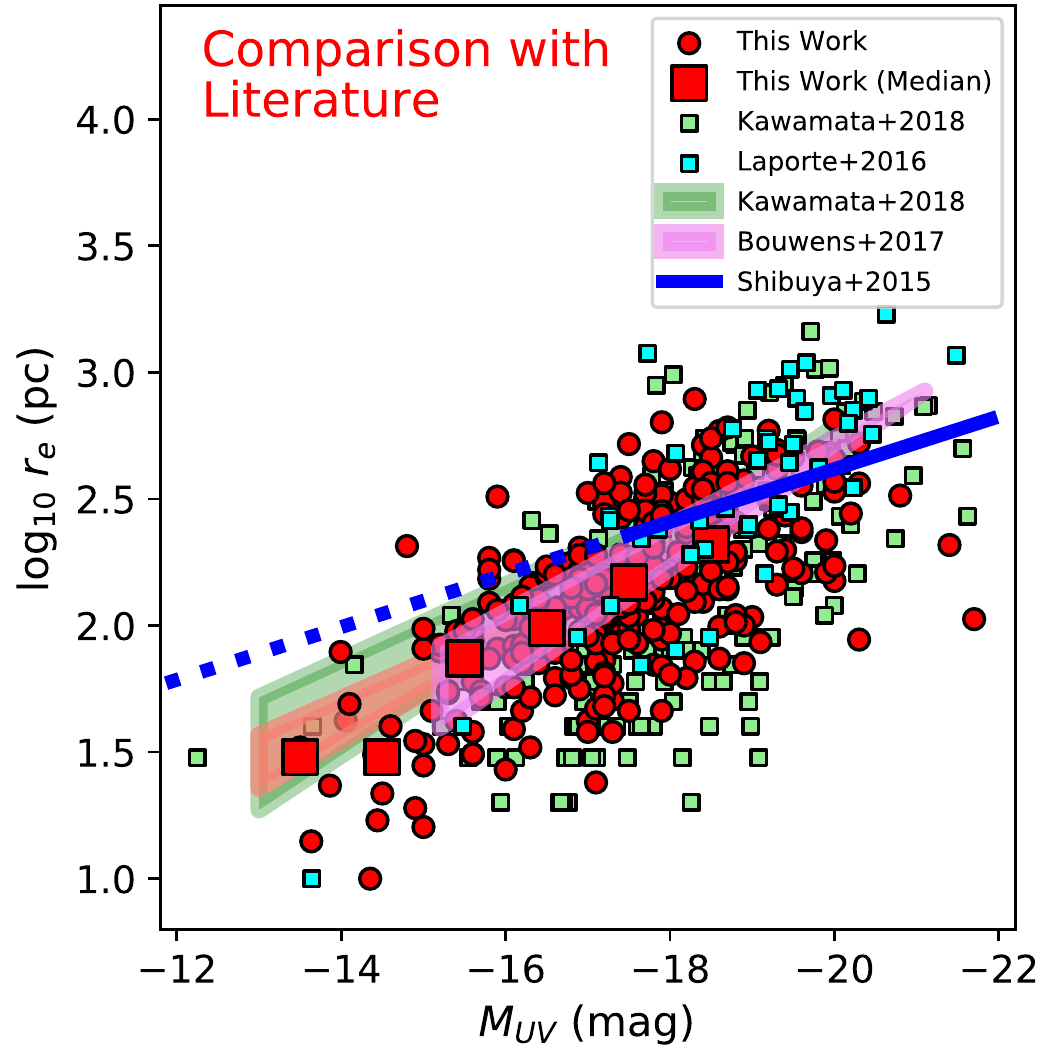}
\caption{Comparison of our size vs. luminosity measurements
  (\textit{red circles}) with earlier measurements by Laporte et
  al.\ (2016: \textit{light open blue circles}) and Kawamata et
  al.\ (2018: \textit{light green open circles}).  Also shown are the
  median sizes derived in this analysis (\textit{red squares}) as a
  function of $UV$ luminosity $M_{UV}$ as well as the size-luminosity
  relations derived by Bouwens et al.\ (2017: \textit{violet line})
  and Kawamata et al.\ (2015) for their $z\sim6$-8 selection.  The
  shaded orange, violet, and green regions indicate the constraints on
  the median sizes vs. $UV$ luminosity derived in this study, Bouwens
  et al.\ (2017a), and Kawamata et al.\ (2018), respectively.  The
  steepness of the size-luminosity relation derived from our
  measurements contrasts strikingly with the shallower slope of the
  Shibuya et al.\ (2015) relation at higher luminosities, and its
  extrapolation to lower luminosities.\label{fig:sizelit7}}
\end{figure}

\begin{figure}
\epsscale{1.15}
\plotone{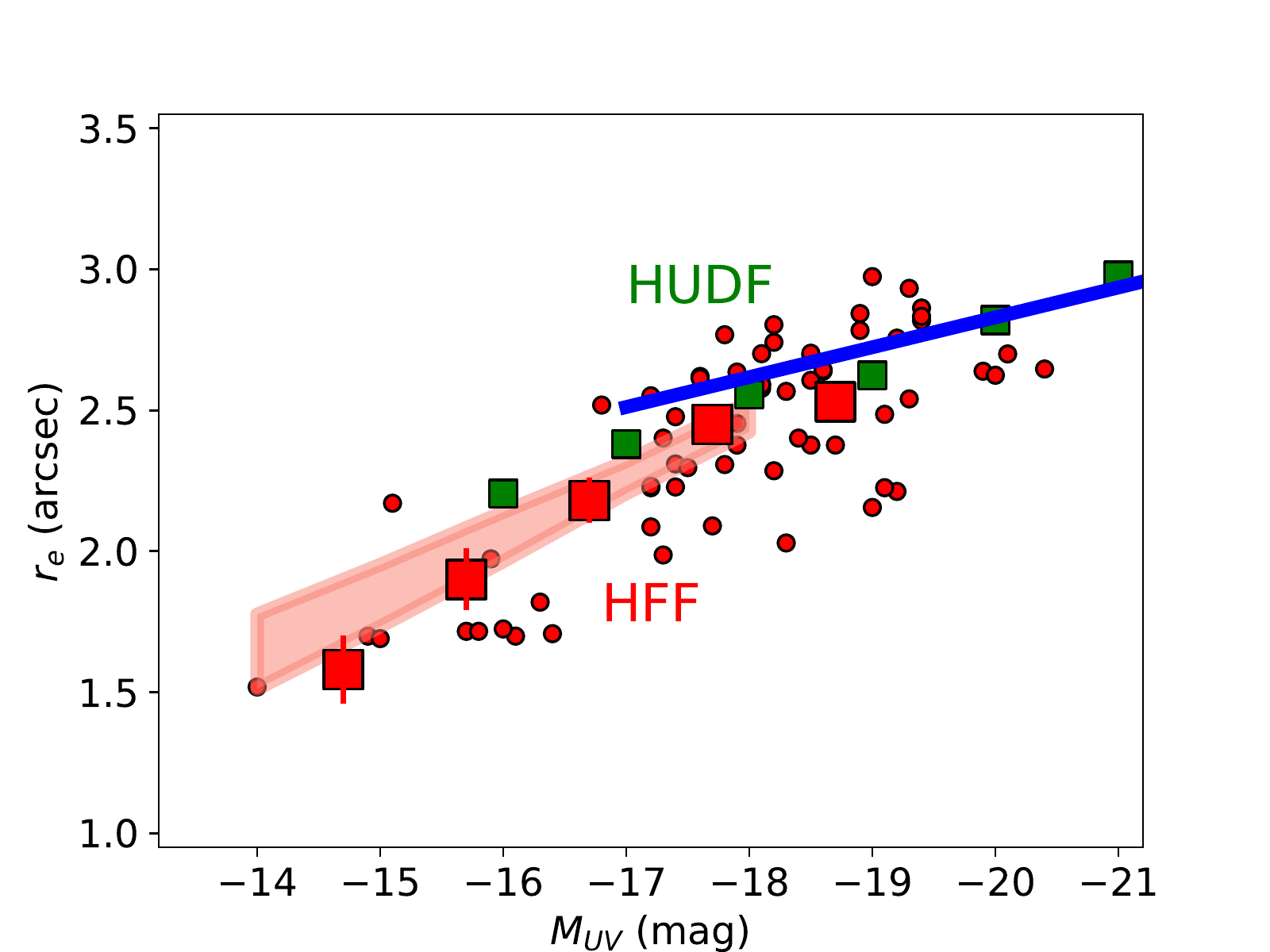}
\caption{Similar to Figure~\ref{fig:sizefit7} but for sources in our
  $z\sim4$ selection.  The green squares show the median sizes we
  measure with \textsc{galfit} for galaxies at $z\sim4$ as a function
  of $UV$ luminosity.  Similar to the situation at $z\sim6$-8, the
  size-luminosity relation for lower luminosity galaxies appears to be
  steeper than at higher luminosities.\label{fig:sizefit4}}
\end{figure}

\begin{figure*}
\epsscale{1.15}
\plotone{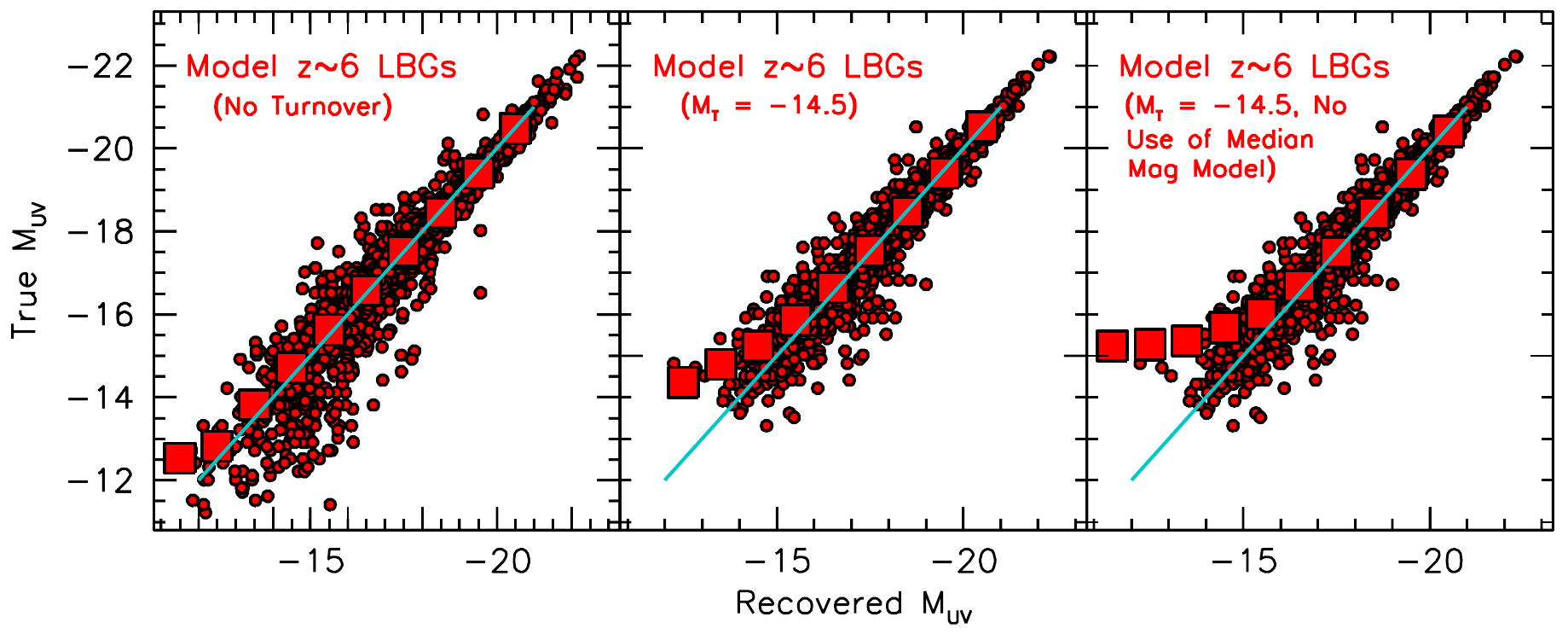}
\caption{``True'' $UV$ luminosity vs. the $UV$ luminosity estimated
  from the median parametric model for sources in a large
  forward-modeling simulation.  The red circles show the original
  model $UV$ luminosities plotted against the recovered $UV$
  luminosities from the median magnification maps in the left and
  center panels and the Keeton magnification map in the right panel.
  The large red squares show the median ``true'' $UV$ luminosity per
  magnitude bin of recovered $UV$ luminosity.  The input LF in the
  left panel has a faint-end slope of $-2.05$ with no turn-over, while
  in the center and right panel, the input LF has a turn-over at
  $-14.5$.  For an input LF with no turnover, the ``true'' $UV$
  luminosities for sources in the faintest luminosity bins is similar
  or only slightly brighter than the nominal luminosities inferred
  from the lensing models.  Meanwhile, for input LFs with a turn-over
  at $-14.5$ mag, the ``true'' luminosity for sources in the faintest
  luminosity bins (i.e., $>$$-$15 mag) are found to be typically
  $\sim$1-3 mag brighter (in the median) than the nominal luminosities
  inferred from the lensing models.  The luminosities recovered using
  the median magnification model are more in line with the ``true''
  luminosities than using magnification factors from a single model
  (as shown in the right panel for the Keeton
  models).\label{fig:muvrec}}
\end{figure*}

\subsection{Size-Luminosity Relation\label{sec:fit}}

Having set up a forward-modeling procedure to simulate a mock
size-luminosity distribution which includes both the necessary
selection effects and magnification uncertainties, we can now compare
the derived size-luminosity distribution with the forward-modeled
results to derive a maximum likelihood size-luminosity relation.

In ascertaining the maximum likelihood size-luminosity relation, we
systematically compared the forward-modeled size-luminosity
distribution to the observed distribution to assess goodness of fit.
For this, we computed both the mean size and scatter in the 1-mag
intervals from $-13.5$ to $-18.5$ mag, compared it with the observed
quantities, and computed a total $\chi^2$.

Then, for the second step, we compared the number of forward-modeled
sources in the three faintest magnitude intervals that made up each
selection with the observed number of sources in each magnitude
interval.  We focused on the three faintest magnitude bins -- since
this is where the size distribution is expected to have a particularly
significant impact on the number of sources selected (e.g., Bouwens et
al.\ 2017a).  For our $z\sim 6$-8 selections, the three faintest
magnitude intervals were $-15.5$, $-14.5$, and $-13.5$ mag,
respectively, while for our $z\sim 4$ selections, the three faintest
magnitude intervals were $-14.5$, $-15.5$, and $-16.5$ mag.

For each of these magnitude intervals, we computed a total $\chi^2$
based on the differences in the total number of sources and added that
to $\chi^2$ we derived comparing the mean size and scatter.  We then
computed a relative likelihood for model size-luminosity relation as
$e^{-\chi^2 / 2}$.

We then adopted a MCMC procedure to derive the maximum likelihood
values for the slope $\alpha$, scatter $\sigma$, and offset $r_0$ to
the size-luminosity relation $r_e = r_0 (L_{UV}/L_0)^{\alpha}$ for
both our $z\sim6$-8 sample and our $z\sim4$ sample.  The best-fit size
luminosity relation we derive for our $z\sim6$-8 sample is $\log_{10}
(r_0/\textrm{pc}) = 2.76 \pm 0.07$, $\alpha = 0.40 \pm 0.04$, and
$\sigma = 0.21 \pm 0.03$ dex.  This relation is shown in
Figure~\ref{fig:sizefit7} as the shaded red region, and it is clearly
steeper ($3\sigma$ significance) than found by Shibuya et al.\ (2015)
at brighter magnitudes where the effective slope to the relation is
$0.26\pm0.03$.

The previous findings for the size-luminosity relationship for
lower-luminosity galaxies found by Bouwens et al.\ (2017) and Kawamata
et al.\ (2018) are consistent with our new and arguably more robust
result.  These two previous studies found the median sizes of galaxies
to depend on the $UV$ luminosity as $L^{0.50\pm0.07}$ and
$L^{0.46_{-0.09}^{+0.08}}$, respectively.  In both cases, the
relations are consistent with what find here, albeit with a slightly
steeper slope.

We show both our own determinations and constraints on the
size-luminosity relation from Bouwens et al.\ (2017) and Kawamata et
al.\ (2018) as the red, violet, and green shaded regions in
Figure~\ref{fig:sizelit7}.  Also included on this figure are
individual size measurements from Kawamata et al.\ (2018: \textit{open
  green circles}) and Laporte et al.\ (2016: \textit{open blue
  circles}), and this work (\textit{small filled red circles}).
Comparing our new more robust constraints on the size-luminosity
relationship at $z\sim6$-8 with previous determinations, we can see
there is a broad consistency between the results.

For our $z\sim4$ sample, the best-fit size luminosity relation we
derive is $\log_{10} (r_0/\textrm{pc}) = 3.13 \pm 0.10$, $\alpha =
0.54 \pm 0.08$, and $\sigma = 0.14 \pm 0.03$ dex.  This relation is
shown in Figure~\ref{fig:sizefit4} as the shaded red region.  While
not as well determined given the smaller number of sources, it is
clearly steeper than the $z\sim4$ relation found at higher
luminosities by Shibuya et al.\ (2015) or as found from the median
sizes of $z\sim4$ galaxies from the HUDF.  In deriving the median
sizes from the HUDF, we made use of our own size measurements using
\textsc{galfit} (Peng et al.\ 2002).

The scatter $\sigma$ we find in the size-luminosity relation at both
$z\sim4$ and $z\sim6$-8, i.e., $\sigma = 0.18\pm 0.04$ and $0.21 \pm
0.03$, is also consistent with the $\sim$0.24 dex-intrinsic scatter
that Shibuya et al.\ (2015) find for more luminous star-forming
galaxies at $z\sim4$-10.

As part of our forward-modeling procedure, a natural output is a
prediction of the number of sources with very small sizes, and the
result is very instructive.  For our fiducial size-luminosity relation
at $z\sim4$ and $z\sim6$-8, we predict 1$\pm$1 and 11$_{-6}^{+9}$
sources, respectively, with sizes $<$40 pc, vs. 3 and 23 sources,
respectively, that we find in the observations to have such small
sizes.  If instead we take the slope of the size-luminosity relation
to be 0.26, as per Shibuya et al.\ (2015), the predicted number of
such small sources is 0.01$\pm$0.01 and 1$_{-1}^{+2}$, respectively.
The observed number of very small sources clearly supports the
existence of a steeper size-luminosity relation.  We discuss the
physical interpretation of the very small sources we find as part of
this study in a companion paper (Bouwens et al.\ 2021).

\subsection{Impact of Lensing Uncertainties on the $UV$ Luminosities Recovered at  $-$15 mag\label{sec:muvrecov}}

Here we ask how well the inferred luminosities of sources behind the
HFF clusters actually track their true luminosities.  Addressing this
question is not simple and requires significant testing through
simulation and recovery experiments.  Previous work included both
model-to-model comparisons (Priewe et al.\ 2017; Bouwens et
al.\ 2017b) and end-to-end tests (Meneghetti et al.\ 2017).  These
studies have demonstrated that lensing models appear to be reasonably
predictive to magnification factors of 30 in the median, but with
0.4-0.5 dex scatter (see e.g. Figure 3 from Bouwens et al.\ 2017b).
\textit{Despite their utility, none of these earlier tests were framed
  in terms of the $UV$ luminosity in particular.}

The purpose of this section is to look at the extent to which sources
identified as having a given $UV$ luminosity actually have that $UV$
luminosity in the median.  Framing the tests in terms of luminosity
(instead of magnification) is valuable since sources from many
different magnification and apparent magnitude bins contribute to a
given bin in $UV$ luminosity and the total volume within various bins
of $UV$ luminosity varies quite dramatically.

To determine how well the inferred $UV$ luminosities actually track
the true $UV$ luminosities, we use the forward-modeling methodology
described in Bouwens et al.\ (2017b).  We use the v4.1 CATS
magnification models to create mock catalogs over the five HFF
clusters where these models are available and in one cluster the v4
CATS magnification model.  The input LF used in constructing mock
catalogs in the simulation has a faint-end slope of $-2.05$ and either
no turn-over or a turn-over at $-14.5$ mag.  The adopted functional
form of the $UV$ LF around the turn-over is as presented in Bouwens et
al.\ (2017b).  Each source is assigned coordinates and an apparent
magnitude.  Absolute magnitudes are derived for sources in these
catalogs based on either the median magnification from the latest
parametric lensing models or the magnification from the Keeton
models.

Both the true $UV$ luminosities and the luminosities recovered from
the median, median, or Keeton models are presented in the left,
center, or right panels of Figure~\ref{fig:muvrec}.  Results in the
left panel are based on an input LF with no turn-over, while those in
the center and right panels are based on an input LF with a turn-over
at $-14.5$ mag.  Also shown with the red squares are how well the
luminosities of sources drived from the median magnification map
predict the ``true'' model luminosities.  Interestingly, the
luminosities of sources inferred from the median magnification map
track the actual model luminosities reasonably well in the case of a
LF with no turn-over, but show offsets relative to the true values
faintward of $-15$ mag for a LF with a turn-over at $-14.5$ mag.
Given uncertainty regarding whether a turn-over in the LF exists
faintward of $-15$ mag (e.g., Atek et al.\ 2018), care may need to be
exercised to quantify both the characteristics and luminosity function
of sources fainter than $-$15 mag using current data sets.

\begin{figure}
\epsscale{1.15}
\plotone{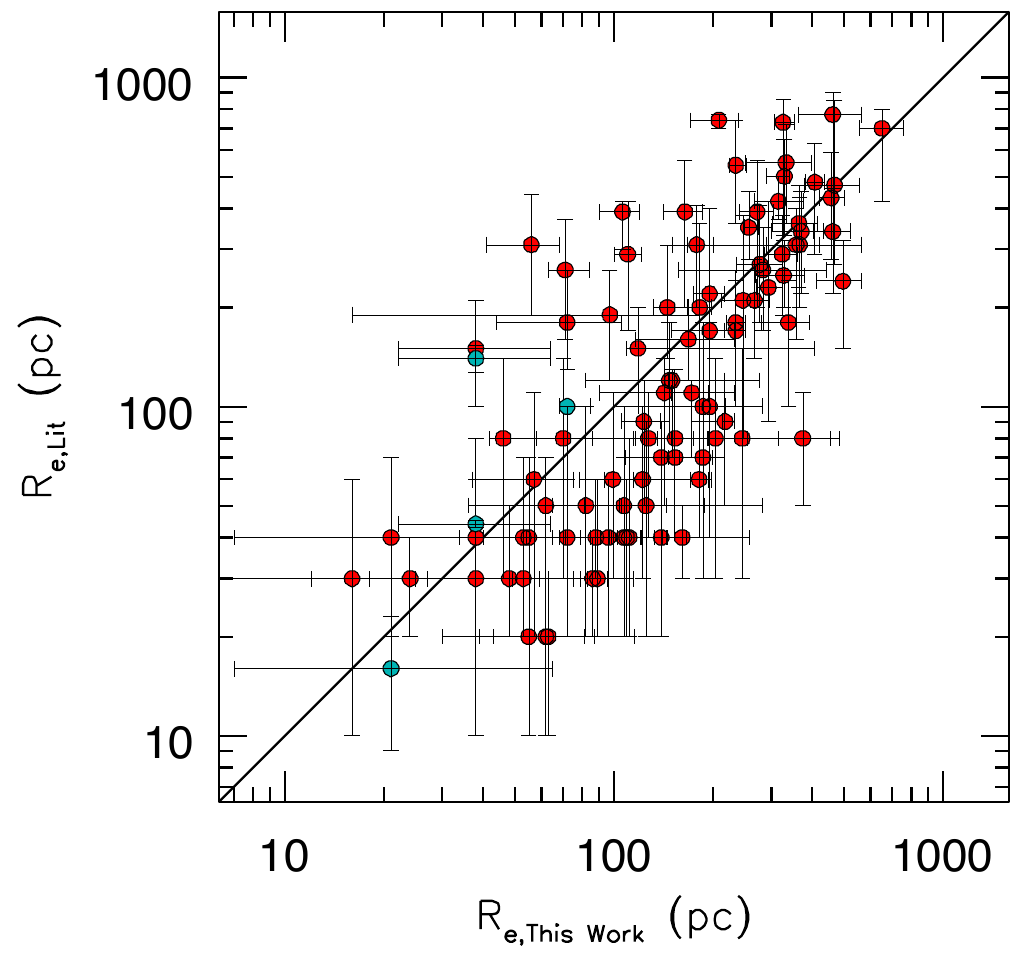}
\caption{Comparison of the lensing-corrected size measurements derived
  here (\textit{horizontal axis}) with those derived in previous
  studies (\textit{vertical axis}).  The red circles represent
  comparisons with the Kawamata et al.\ (2018) measurements and the
  cyan circles represent comparisons with Vanzella et al.\ (2017a) and
  Vanzella et al.\ (2019).  For individual sources, we find a typical
  0.3-dex difference in the measured sizes, but for the median source,
  our size measurements are in excellent agreement with previous work.
\label{fig:litcomp}}
\end{figure}

\section{Discussion}

\subsection{Comparison with Size Measurements in Previous Work}

Before discussing the form of the size-luminosity relation at high
redshift, we compare our results against size measurements made by
other teams on the same sources that make up our samples.

Of particular interest are new results recently obtained by Kawamata
et al.\ (2018), who have updated the results from Kawamata et
al.\ (2015) to include sources from all six HFF clusters and parallel
fields.  Cross-matching our source catalogs with sources in the
Ishigaki et al.\ (2017)/Kawamata et al.\ (2018) catalogs, we find 92
sources in common.  Our size measurements agree fairly well with those
from Kawamata et al.\ (2018) in terms of the median sizes, with our
measurements being 25$\pm$7\% larger, but with a $1\sigma$ scatter of
0.30 dex in the measured sizes of individual sources
(Figure~\ref{fig:litcomp}).  The differences likely reflect
source-to-source differences in the magnification factors Kawamata et
al.\ (2018) infer from the \textsc{GLAFIC} models they utilize and the
median models we utilize.  The significant source-to-source scatter in
the measured sizes suggests there are large model-dependent
uncertainties in establishing the sizes of individual galaxies.

We can also see how well our respective size measurements agree for
sources with particularly small sizes.  For sources where we estimate
sizes less than 50 pc, the median size measurement in their catalog is
35 pc.  Similarly, when Kawamata et al.\ (2018) estimate sizes less
than 50 pc, the median size measurement in our catalog is 63 pc.  As
such, there is reasonable agreement (at least in the median) between
our estimated sizes and those of Kawamata et al.\ (2018) and also our
selected samples of sources with small sizes and those of Kawamata et
al.\ (2018).

We also compare our size measurements to those made by Vanzella et
al.\ (2017a, 2019) on a few particularly compact sources.  Vanzella et
al.\ (2017a, 2019) examined these sources due to their particularly
small sizes and proximity to the star cluster regime.  For Vanzella et
al.\ (2017a) sources GC1, D1, and D2, we measure half-light radii of
21$_{-14}^{+44}$ pc, 38$_{-16}^{+26}$ pc, and 72$_{-4}^{+13}$ pc
vs. measurements of 16$\pm$7 pc, 140$\pm$13 pc, and $<$100 pc,
respectively, from Vanzella et al.\ (2017a).  In a follow-up study,
Vanzella et al.\ (2019) reanalyzed the size and structure of their D1
source, finding a size of $<$13 pc for the core region and a size of
44 pc for the galaxy as a whole.  For GC1 and D2, our measurements
agree with the Vanzella et al. (2017a) measurements within the
$1\sigma$ uncertainties, while for D1, our measurements agree with the
later Vanzella et al.\ (2019) measurements.

In summary, we find that there is reasonable agreement in the median
between our size measurements and that obtained in earlier work
(Kawamata et al.\ 2018; Vanzella et al.\ 2017a, 2019).  This is
encouraging, given the expected scatter in the measured sizes for
individual sources resulting from our reliance on different lensing
models (i.e. a median model) than Kawamata et al.\ (2018) and Vanzella
et al.\ (2017a, 2019) utilize.

\begin{figure}
\epsscale{1.17}
\plotone{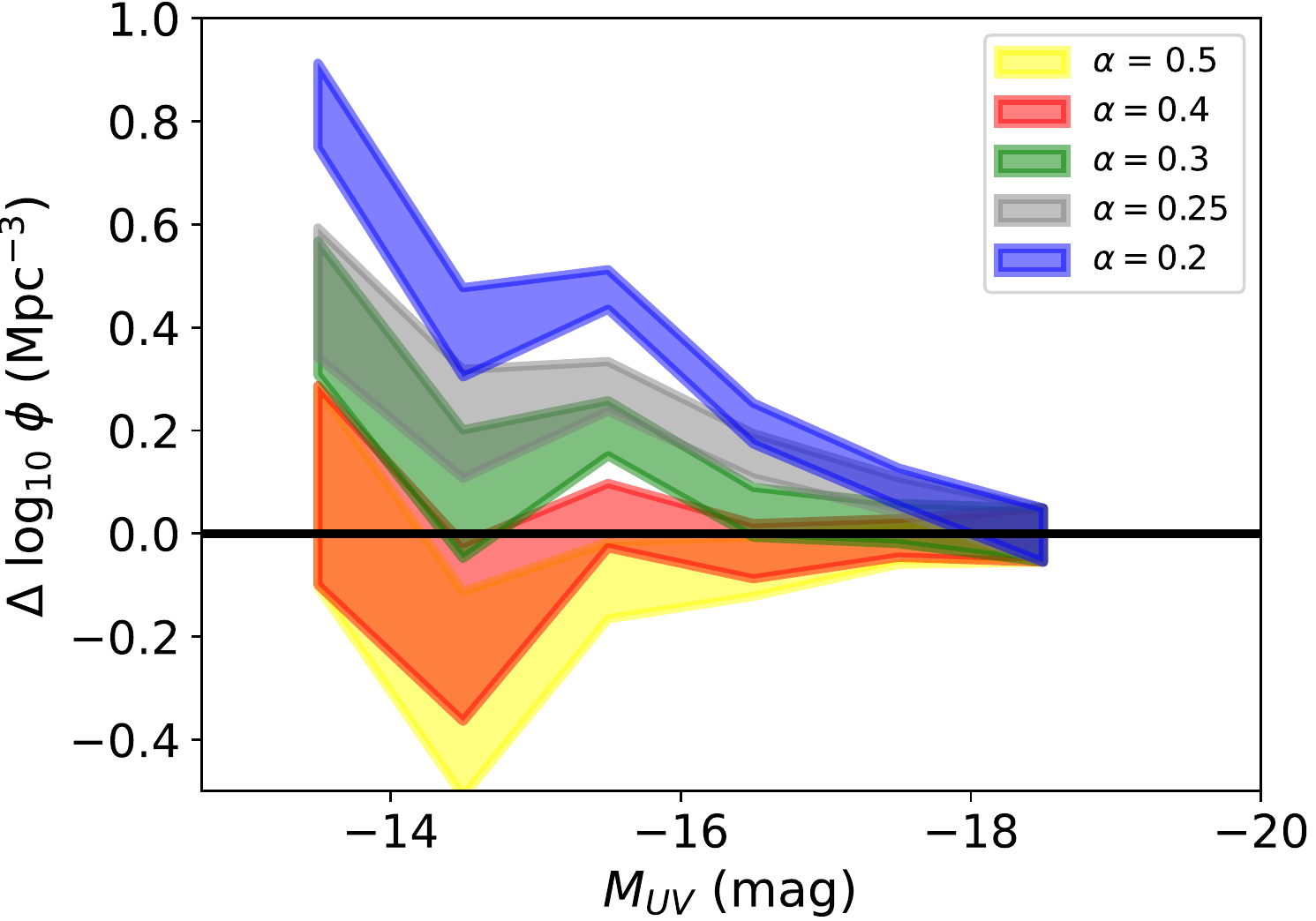}
\caption{The impact of the completeness at the faint end of $z\sim6$-8
  selections on the inferred $UV$ LF.  Shown with the shaded yellow,
  red, green, gray, and blue regions is the excess in the inferred
  $UV$ LF assuming a true size-luminosity scalings of $r\propto
  L^{0.4}$, but recovering the results assuming a size-luminosity
  scaling of $L^{0.5}$, $L^{0.4}$, $L^{0.3}$, $L^{0.26}$, and
  $L^{0.2}$, respectively.  Results for the Shibuya et al.\ (2015)
  size luminosity relation are shown with gray region.  The plotted
  regions enclose the $1\sigma$ Poissonian errors on the $UV$ LF
  results based on the number of sources in each absolute magnitude
  interval of our $z\sim6$-8 selections.  The assumption of shallow
  size-luminosity relation can substantially steepen the inferred
  faint-end slope $\alpha$ for the $UV$ LF and also cause the $UV$ LF
  to show a potential upturn (Figure~\ref{fig:lfshape}).\label{fig:upturn}}
\end{figure}

\begin{figure*}
\epsscale{1.17}
\plotone{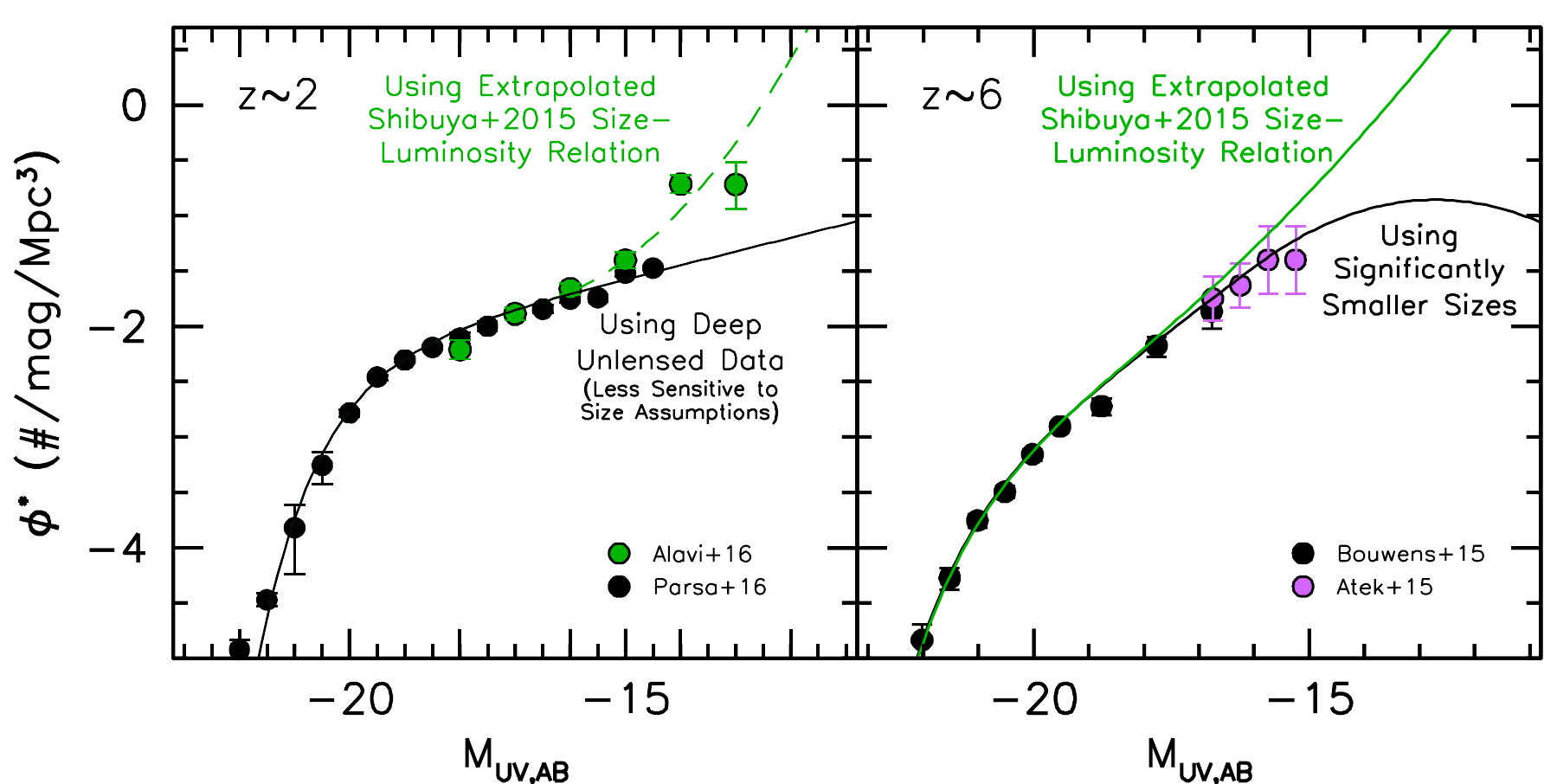}
\caption{An illustration of the significant impact that high levels of
  incompleteness can have on the faint-end ($>$$-$15 mag) form of
  $z=2$-8 LFs.  Given that the incompleteness is directly calculable
  from the assumed size-luminosity relation, the faint-end form of the
  LF has a direct impact on the inferred size distribution of faint
  galaxies.  If faint $z\sim2$ and $z\sim6$ galaxies have sizes which
  are a simple extrapolation of the Shibuya et al.\ (2015)
  size-luminosity relation, the recovered $UV$ LFs at $z\sim2$ and
  $z\sim6$ combining blank field and lensing cluster observations are
  as indicated by the green lines and points (Alavi et al.\ 2016;
  \S5.4 of Bouwens et al.\ 2017b).  Meanwhile, the recovered UV LFs
  show much lower volume densities if faint galaxies have
  significantly smaller sizes than inferred from an extrapolation of
  the Shibuya et al.\ (2015) size-luminosity relation, as we have
  discussed previously (so leading to a break in this relation -- as
  demonstrated nicely in Figure~\ref{fig:sizefit7}).  The black line in the
  right panel is the Bouwens et al.\ (2017b) LF result and relies on
  the higher detectability of faint sources expected from the
  size-luminosity relationship derived here (where sources are smaller
  than extrapolating the Shibuya et al.\ 2015 relation).  The right
  panel also shows the blank field $z\sim6$ LF results from Bouwens et
  al.\ (2015: \textit{black solid circles}) along with the results of
  Atek et al.\ (2015: \textit{magenta solid circles}).  The black line
  in the left panel is from Parsa et al.\ (2016) and derived from the
  sensitive blank field observations over the XDF/HUDF (\textit{black
    line and black circles}).  In this case, size assumptions are not
  especially important at the faint end since sources are smaller than
  the PSF.  If we suppose -- following most theoretical models -- that
  the $UV$ LF at $z\sim2$ and $z\sim6$ extends towards fainter
  luminosities with a fixed (or progressively flatter) faint end
  slope, then the size-luminosity relation cannot extend to the lowest
  luminosity galaxies following the Shibuya et al.\ (2015) scaling,
  but must show a break at some luminosity towards a steeper scaling.
  While circumstantial, the present argument is very strong for
  fainter galaxies having small sizes and the size-luminosity relation
  showing a break at $\sim$$-$17 mag.\label{fig:lfshape}}
\end{figure*}

\subsection{Implications for Sizes from the Faint End Form of the $z\sim6$ LF Derived from the HFFs}

As discussed in much previous work (e.g., Grazian et al.\ 2011;
Bouwens et al.\ 2017a, 2017b; Kawamata et al.\ 2018; Atek et
al.\ 2018), there is a direct connection between (1) the distribution
of sizes and surface brightnesses assumed for the lowest luminosity
galaxies and (2) the faint-end form inferred for the $UV$ LFs at
$z\sim2$-6.  The purpose of this section is to spell out this
important connection and the impact one has on the other.

\subsubsection{Implications of Standard Shallow Size-Luminosity Relations for the Faint-End Form of the $UV$ LFs}

As we previously discussed in \S3 above, blank-field studies have
found that the median half-light radius of brighter galaxies depends
on the luminosity $L$ of galaxies as $R\propto L^{0.26}$ (Shibuya et
al.\ 2015), across a wide range of redshifts.  Huang et al.\ (2013)
find a similar scaling at $z\sim4$ and $z\sim5$, and we might expect
similar scalings to apply to higher redshift galaxies if we
extrapolate the size-mass relations obtained by van der Wel et
al.\ (2014) and Mowla et al.\ (2019a).

\begin{figure*}
\epsscale{1.15}
\plotone{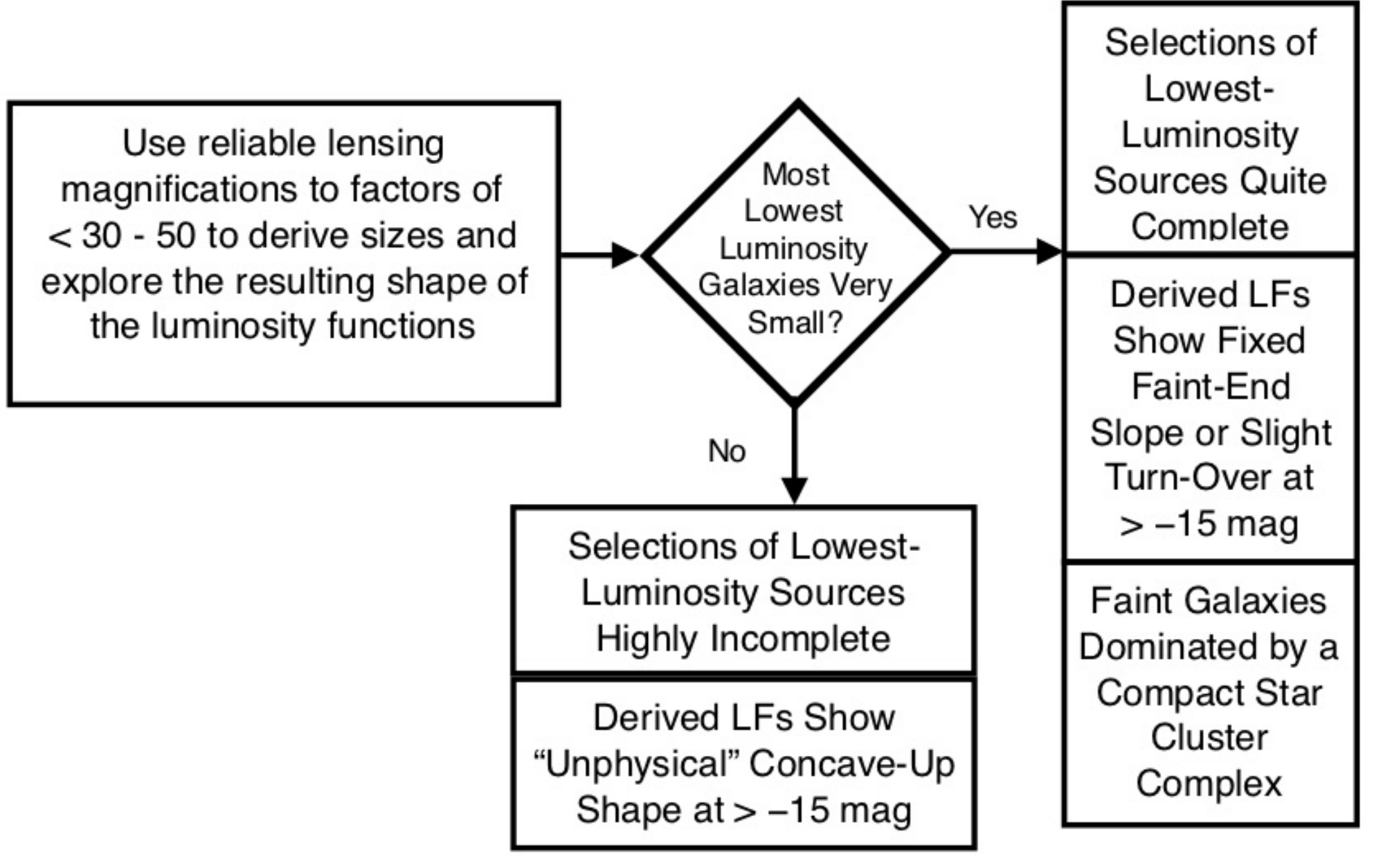}
\caption{A simple flowchart summarizing the connection between the
  form of the $UV$ LF at high-redshift and the implied size
  distribution for lower luminosity galaxies (see \S5.3).  An
  essential starting point for probing LFs and sizes to very low
  luminosities is the assumption that the HFF lensing models are
  predictive to magnification factors of $>$10 (e.g., as the
  Meneghetti et al.\ 2017; Priewe et al.\ 2017; Bouwens et al.\ 2017b
  results suggest, and as we also discuss in \S\ref{sec:reliable}) to
  make any inferences.  If we assume that the lower-luminosity
  galaxies have sizes that are simply follow an extrapolation of the
  Shibuya et al.\ (2015) size-luminosity relation (where $r\propto
  L^{0.26}$), this implies a $UV$ LF with concave-upwards form at
  $>$$-$15 mag (see Figure~\ref{fig:lfshape}), which would be
  inconsistent with all theoretical models.  On the other hand, if one
  supposes one should recover a standard faint-end form for the $UV$
  LF, one must assume a steep size-luminosity relation, e.g.,
  $r\propto L^{0.5}$.  \textit{The observations do not allow for the
    assumption of both (1) a conventional size-luminosity relation
    (with $r\propto L^{0.26}$) and (2) a conventional faint-end form
    for the $UV$ LF at $>$$-$15 mag.}\label{fig:flowchart}}
\end{figure*}

If these scalings apply to extremely low luminosity $z\sim6$-8
galaxies, the surface brightness should vary as $L/R^2 \propto
L^{0.5}$.  With such a scaling, 0.001$L^*$ ($-$13.5 mag) galaxies
would have surface brightnesses 30$\times$ fainter than $L^*$ galaxies
have.  At such low surface brightnesses, we would expect searches for
faint $z=2$-8 galaxies to be highly incomplete.  

If we apply corrections for the expected high incompleteness in high
magnification regions (from the extrapolated Shibuya et al.\ 2015
relation) to the present $z\sim6$-8 search results and Ishigaki et al.
(2018: their Figure 1) -- who find virtually the same surface density
of $z\sim6$ candidate galaxies in both high and low magnification
regions -- we would infer very high volume densities for the ultra-low
luminosity sources at $z\sim6$.  The gray-shaded region in
Figure~\ref{fig:upturn} shows the excess in the volume density of
lower luminosity galaxies we would derive if the extrapolated Shibuya
et al. (2015) relation for estimating the incompleteness relative to
the fiducial $r\propto L^{0.4}$ relation derived here.  The blue,
green, red, and yellow shaded regions in Figure~\ref{fig:upturn} show
the excess that would be inferred for other size-luminosity scalings,
assuming $r \propto L^{0.2}$, $r \propto L^{0.3}$, $r \propto
L^{0.4}$, and $r \propto L^{0.5}$, respectively.

While one impact of the shallower size-luminosity relations is to
steepen the faint-end slope of the $UV$ LF -- with an impact of
$\Delta \alpha = -0.05\pm0.04$, 0.20$\pm$0.04, 0.28$\pm$0.03, and
0.42$\pm$0.03 on the faint-end slope $\alpha$ for $r\propto L^{0.5}$,
$r\propto L^{0.3}$, $r\propto L^{0.26}$, and $r\propto L^{0.2}$,
respectively -- changes to the size-luminosity relations also cause
the $UV$ LF to increasingly take on a concave-upwards form.  The
approximate impact of this adopting the Shibuya et al.\ (2015)
size-luminosity relation is illustrated with the dashed green line in
the right panel of Figure~\ref{fig:lfshape}.  Given the $1\sigma$
uncertainties on the faint-end slope at $z\sim6$ of 0.08, a
concave-upwards form could be readily seen (at $2\sigma$ significance)
in size-luminosity relations as shallow as $L^{0.3}$ (where the impact
on the effective $\alpha$ at the faint-end of the LF would be
0.20$\pm$0.04).

Earlier, applying an extrapolation of the size-luminosity relation
obtained by Shibuya et al.\ (2015) for $z\sim2$ galaxies -- with a
similar size-luminosity dependence to their $z\sim6$ results -- Alavi
et al.\ (2016) had derived a $UV$ LF at $z\sim2$ showing exactly such
a concave-upward form.\footnote{Alavi et al.\ (2016) explicitly looked
  at the dependence of the volume density of faint sources on the
  assumed source size.  While they find a dependence on size, they
  argue against the extreme faint end of the LFs having an upturn.
  Instead, they argue for a consistent faint-end slope of their LF
  over the full luminosity range they consider, from $-19$ to $-15$
  and also faintward of $-$15 mag.} This is indicated with the green
solid line in the left panel of Figure~\ref{fig:lfshape}.

For the situation where no luminosity dependence is assumed to the
size distribution, an even steeper concave upwards form would be
expected than the already dramatic upturns shown with the green lines
in Figure~\ref{fig:lfshape}.  This is true even assuming sizes as
small as 0.4 kpc (as is observed at $-18$ mag: e.g., Ono et al.\ 2013;
Shibuya et al.\ 2015) or even assuming a broad range in widths to the
size distribution.  For such size distributions, the recovered $UV$
LFs we find based on our own simulations (see also Kawamata et
al.\ 2018) would be much higher in terms of their volume densities
than what has generally been reported in the literature at $>$$-$15
mag (e.g., Livermore et al.\ 2017; Bouwens et al.\ 2017b; Ishigaki et
al.\ 2018; Atek et al.\ 2018; Yue et al.\ 2018).

Similar to the analysis provided by Bouwens et al.\ (2017b) in their
\S5.4, Atek et al.\ (2015) and Castellano et al.\ (2016) made use of
standard shallow size-luminosity relations in deriving the LF at
$z\sim6$-7.  These studies found that the only way they could obtain
plausible LF results was through the restriction of their
determinations to sources brightward of $-$15 mag.  H. Atek (private
communication) indicated to us that they did not extend their LF
results faintward of $-$15 mag, due to uncertainties in extrapolating
the size-luminosity relation into this regime and the very high volume
densities implied at such faint magnitudes by the uncertain
incompleteness corrections.

\subsubsection{Possibility of a Steep Size-Luminosity Relation?}

While one would expect to derive a ``concave-upwards'' luminosity
function (see e.g. Figure~\ref{fig:lfshape}) for galaxies at $z\sim6$
making use of the standard shallow size-luminosity relation for
completeness measures, there are observational and theoretical reasons
for disfavoring such a ``concave-upwards'' luminosity function.  The
lack of an apparent upturn in the observed faint LF at low redshift
already sets constraints on the existence of such an upturn.
Additionally, as demonstrated by Weisz et al.\ (2014) and
Boylan-Kolchin et al.\ (2014, 2015), abundance matching of nearby
dwarf galaxies sets an upper limit on the volume density of lower
luminosity galaxies in the high-redshift universe.\footnote{One
complication, of course, for such constraints is the likely
incompleteness in low-redshift samples and large scatter in the
stellar-mass halo-mass relation.}

From a theoretical perspective, one would expect the faint-end of the
LF to continue to largely trace the halo mass function, but at the
extreme faint end, the $UV$ LF is expected to flatten or even turn
over, as a result of increasingly inefficient gas cooling and
radiative heating.  A typical turn-over luminosity is $\sim-12$ mag
(Liu et al.\ 2017; Finlator et al.\ 2015; Gnedin 2016; O'Shea et
al.\ 2015; Ocvirk et al.\ 2016; Yue et al.\ 2016; Dayal et al.\ 2014).
Theoretical LFs are thus not expected to become steeper towards the
extreme low luminosity end.

If we discount such an upward change in the slope on the basis of
these plausibility arguments, we must necessarily assume that the
size-luminosity relation must show a break at $\sim$$-$16 to
$\sim$$-$17 mag, such that lower luminosity galaxies are all very
small and have high surface brightnesses.  This would translate to
generally high levels of completeness in searches for lower luminosity
galaxies to high magnification factors, as we in fact make use of as
part of the present analysis.  Indeed, the most natural way to explain
the constant surface density of $z\sim6$ candidate galaxies found by
both Bouwens et al.\ (2017b) and Ishigaki et al.\ (2018) is to suppose
that the completeness of $z\sim6$ selections remain high even to high
magnification factors.  Bouwens et al.\ (2017b) made use of such small
size assumptions in deriving constraints on $z\sim6$ $UV$ LF, finding
a roughly fixed faint-end slope to very low luminosities $>$$-$14 mag,
with a possible turn-over at the faint end.  The best-fit $z\sim6$ LF
results of Bouwens et al.\ (2017b) are shown in the right panel of
Figure~\ref{fig:lfshape} with a black line.

In Bouwens et al.\ (2017a), we presented other independent evidence
suggesting that lower luminosity galaxies have very small sizes.  In
drawing these conclusions, Bouwens et al.\ (2017a) made use of the
impact of lensing shear on the detectability of $z\sim6$-8 sources in
high magnification regions behind lensing clusters.  Critically,
Bouwens et al.\ (2017a) found approximately the surface density of
galaxies in high magnification regions at both low, intermediate, and
high shear factors.  This is significant since fainter sources with
larger sizes, i.e., $>$160-240 pc, would tend to become undetectable
in regions with higher shear.  The fact that $z\sim6$-8 sources are
then found in higher shear regions with a consistent surface densities
to lower shear, high magnification regions provides us with another
argument that fainter ($\gtrsim$$-$16 mag) galaxies are predominantly
small.

As a caveat to this discussion, we should emphasize that the
conclusions that we have drawn in this subsection are sensitive to the
predictive power of the lensing models.  If the lensing models lose
their predictive power above magnification factors of $\sim$10, the
sources that make up our nominally lowest luminosity samples (i.e.,
$M_{UV}>-15$ mag or $M_{UV}>-14$ mag) would instead be prominently
made up of sources at higher intrinsic luminosities, i.e.,
$M_{UV}\sim-15$ mag, scattering to lower lower luminosities due to
uncertainties in the lensing models.  There is, however, evidence that
lensing models (especially the median model) maintain their predictive
power to magnification factors of 30-50, assuming the tests run by
Meneghetti et al.\ (2017), Priewe et al.\ (2017), and Bouwens et
al.\ (2017b) are sufficiently robust to encompass the relevant issues
(and as we also discuss in \S\ref{sec:reliable}).

The arguments presented in this section regarding the form of the
luminosity function are involved, so we summarize them in
Figure~\ref{fig:flowchart} for clarity.  The use of sizes resulting
from a simple extrapolation of the size-luminosity relation found for
higher luminosity galaxies would suggest very large completeness
corrections and imply an unexpected and rather dramatic upturn.  If
the luminosity function does not show an upturn at very low
luminosities (i.e., $>$$-$15 mag), as predicted in most theoretical
models (e.g., Dayal et al.\ 2014; Gnedin 2016; Liu et al.\ 2016), then
faint sources must be small.

While indirect and relying on theoretical expectations for a
“reasonable” faint end form of the LF, the present argument is
grounded in a robust physical framework. A substantial upturn at the
faint end of the LF, with corresponding larger sizes for the the
detectable regions of galaxies at lower luminosities, would have major
(and unexpected) theoretical implications, while also being in tension
with stellar population analyses of the local dwarf galaxy population
(Weisz et al.\ 2014; Boylan-Kolchin et al.\ 2014, 2015).

The above arguments only add further weight to the direct evidence
presented in \S4 for a steeper size-luminosity relation for galaxies
fainter than $M_{UV} \sim -17$ mag.  Independent evidence for very
small sizes for faint $z\geq4$ galaxies was presented in Bouwens et
al.\ (2017a), who found little change in the prevalence of faint high
redshift galaxies as a function of shear in the lensing field.  As
Bouwens et al.\ (2017a) argue on the basis of extensive simulations,
this can only be the case if faint galaxies are small.

We emphasize that these arguments only apply to the apparent sizes of
sources in the rest-$UV$, which would reveal only the highest surface
brightness star-forming regions (e.g., Overzier et al.\ 2008; Ma et
al.\ 2018), and that the true physical sizes of faint $z\sim4$-8
galaxies may be much larger.  We discuss this situation in more detail
in \S5.1 of a companion paper (Bouwens et al.\ 2021).

\section{Summary}

Here we make use of the unique depth and resolving power of the HFF
cluster observations to examine the sizes and luminosities of 68
$z\sim4$, 184 $z\sim6$, 93 $z\sim7$, and 53 $z\sim8$ sources
identified in the early universe behind the six HFF clusters (68 $z=4$
and 330 $z=6$-8 galaxies in total).

The depth of the HFF observations and the lensing from the massive
foreground clusters make it possible for us to measure the sizes for
$\sim-18$ mag and $\sim-15$ mag galaxies to a typical $1\sigma$
accuracy of $\sim$50 pc and $\sim$20 pc, respectively.  Achieving such
high accuracy on size measurements is crucial for better ascertaining
the physical characteristics of faint star-forming sources in the
early universe.

To obtain the most robust measurements on the sizes and luminosities
of sources, we make use of a MCMC procedure to fit the available
imaging data for each source (\S\ref{sec:measurement}).  We also
utilize the median magnification and shear factors derived from six
different varieties of parametric lensing models CATS, Sharon/Johnson,
GLAFIC, Zitrin-NFW, Keeton, and Caminha.  The model profile is lensed
according to the median magnification and shear factor, convolved with
the PSF, and then compared with a stack of the available WFC3/IR data
on each source.

We show that the majority of our sizes and luminosities should be
reliably measured.  As in our previous work (Bouwens et al. 2017a) and
in other work identifying especially compact sources in the distant
universe (Vanzella et al. 2017a, 2017b, 2019, 2020; Johnson et
al. 2017), the present conclusions do depend on the linear
magnification factors from the lensing models being reliable to
relatively high values of the magnification factor.  In
\S\ref{sec:reliable}, we show that the parametric models should be
predictive to linear magnification factors of $\sim$30.

To derive the size-luminosity relation for star-forming galaxies at
$z\sim4$ and $z\sim6$-8, we need to cope with the impact of surface
brightness selection effects (e.g., Bouwens et al.\ 2017a; Ma et
al.\ 2018) and uncertainties in the magnification model.  Both of
these effects can cause the size-luminosity relation to show an
$r\propto L^{0.5}$ relationship (see e.g. Figure~\ref{fig:lenserror})
at the faint end, which might be steeper than in reality.  We use a
forward-modeling procedure to include these effects in comparing with
the recovered size-luminosity distribution.  In comparing the observed
and forward-modeled distribution, we look at the mean size and scatter
in 1-mag UV luminosity bins and also compare the observed and expected
number of sources in each magnitude bin.

The measured sizes of lensed galaxies in our $z\sim4$ and $z\sim6$-8
samples trend with $UV$ luminosity $L$ approximately as
$L^{0.54\pm0.08}$ and $L^{0.40\pm0.04}$ (Figure~\ref{fig:msre0}: see
\S\ref{sec:fit}) at $>$$-$17 mag.  This is steeper at $3\sigma$ than
the trend found for unlensed luminous ($<$$-$18 mag) sources in the
field, i.e., $r\propto L^{0.26\pm0.03}$ from e.g. Shibuya et
al.\ (2015), suggesting a break in the relation at $\sim$$-$18 mag.
Thanks to our use of a forward-modeling approach, the trends we
recover should be fairly robust and not the result of the
observational effects discussed above which drive a $L^{0.5}$ trend.

Included as part of our forward-modeling fitting results and
supportive of a steeper size-luminosity relation is the surface
density of galaxies in the highest magnification regions behind
clusters and the faint-end form of the $UV$ LF.  To recover $UV$ LF
results where the LF continues with a steep form to faint magnitudes
or only turns over slightly -- similar to that found in theoretical
models (e.g., Liu et al.\ 2016; Gnedin 2016), it is necessary to
assume that the lowest luminosity sources have small sizes (and hence
minimal completeness corrections).  However, under the assumption of a
much shallower size-luminosity relation -- as found by Shibuya et
al.\ (2015) or Mowla et al.\ (2019a) for lower-redshift star-forming
galaxies and the brightest high-redshift sources -- high-redshift
selections would become appreciably incomplete at $>$$-$15 mag.
Applying this incompleteness to the observed surface densities of
faint star-forming galaxies at $z\gtrsim 2$ results in \textbf{an}
unphysical, inferred LF with a concave-upwards form at $>$$-$15 mag
(see \S5.3).  A number of theoretical and observational results argue
against such a concave-upwards form.  This situation is summarized in
Figures~\ref{fig:lfshape} and \ref{fig:flowchart}.

Further support for a steep size-luminosity relation for faint
$z\sim6$-8 galaxies is provided by a similarly steep relationship for
faint $z\sim4$ galaxies and the following indirect evidence: (1) no
clear change in the prevalence of faint galaxies in high magnification
regions as a function of shear (Bouwens et al.\ 2017a), and (2) robust
theoretical expectations for the flattening and turnover of the LF at
very low luminosities (Figure~\ref{fig:flowchart}, \S5.3, Bouwens et
al.\ 2017b).  There is thus significant evidence to support the
size-luminosity relation having a distinctly steeper form, i.e.,
$L^{0.5}$, at lower luminosities, even if the direct measurements we
have at present are not definitive.

In the future, we plan to extend the present analysis by looking at
the sizes and luminosities of star-forming sources at $z=1$-5 behind
the HFF clusters (B. Riberio et al.\ 2021, in prep).  Compact
star-forming sources identified behind the HFFs represent compelling
targets for spectroscopy with both MUSE and JWST to gain more insight
into the nature of these sources.  Among other things, JWST will allow
us to probe the present population to even lower luminosities and with
high levels of completeness, allowing us to achieve an even more
complete characterization of the population of extremely low
luminosity galaxies.

\acknowledgements

We acknowledge stimulating discussions with Xiangcheng Ma.  This work
utilizes gravitational lensing models prodcued by PIs Brada{\v c},
Natarajan \& Kneib (CATS), Merten \& Zitrin, Sharon, and Williams, and
the GLAFIC and Diego groups. This lens modeling was partially funded
by the HST Frontier Fields program conducted by STScI. STScI is
operated by the Association of Universities for Research in Astronomy,
Inc. under NASA contract NAS 5-26555. The lens models were obtained
from the Mikulski Archive for Space Telescopes (MAST).  We acknowledge
the support of NASA grants HST-AR-13252, HST-GO-13872, HST-GO-13792,
and NWO grants 600.065.140.11N211 (vrij competitie) and TOP grant
TOP1.16.057.  PAO acknowledges support from the Swiss National Science
Foundation through the SNSF Professorship grant 190079.  The Cosmic
Dawn Center (DAWN) is funded by the Danish National Research
Foundation under grant No.\ 140.

\appendix

\section{A.  The Completeness of Our HFF $z\geq4$ Selections Shown Using Linear Scalings}

An important aspect of quantifying the characteristics of any
selection is the completeness of that selection, and such is also the
case for the $z\geq 4$ samples presented here over the HFFs.  

As shown in the text the relative completeness of our $z\sim6$-8 and
$z\sim4$ selections depends significantly on both the size and
luminosity of sources.  The relative completeness is shown in
Figures~\ref{fig:msre0} and \ref{fig:msre4} using a logarithmic
scaling.

While a logarithmic scaling can be useful for characterizing the
size-luminosity distribution as a whole relative to the particularly
incomplete regions, i.e., $<$1\%, Figures~\ref{fig:msre0} and
\ref{fig:msre4} are less useful for characterizing the completeness
relative to more modest levels of completeness, i.e., 20\%.

\begin{figure*}
\epsscale{1.15}
\plotone{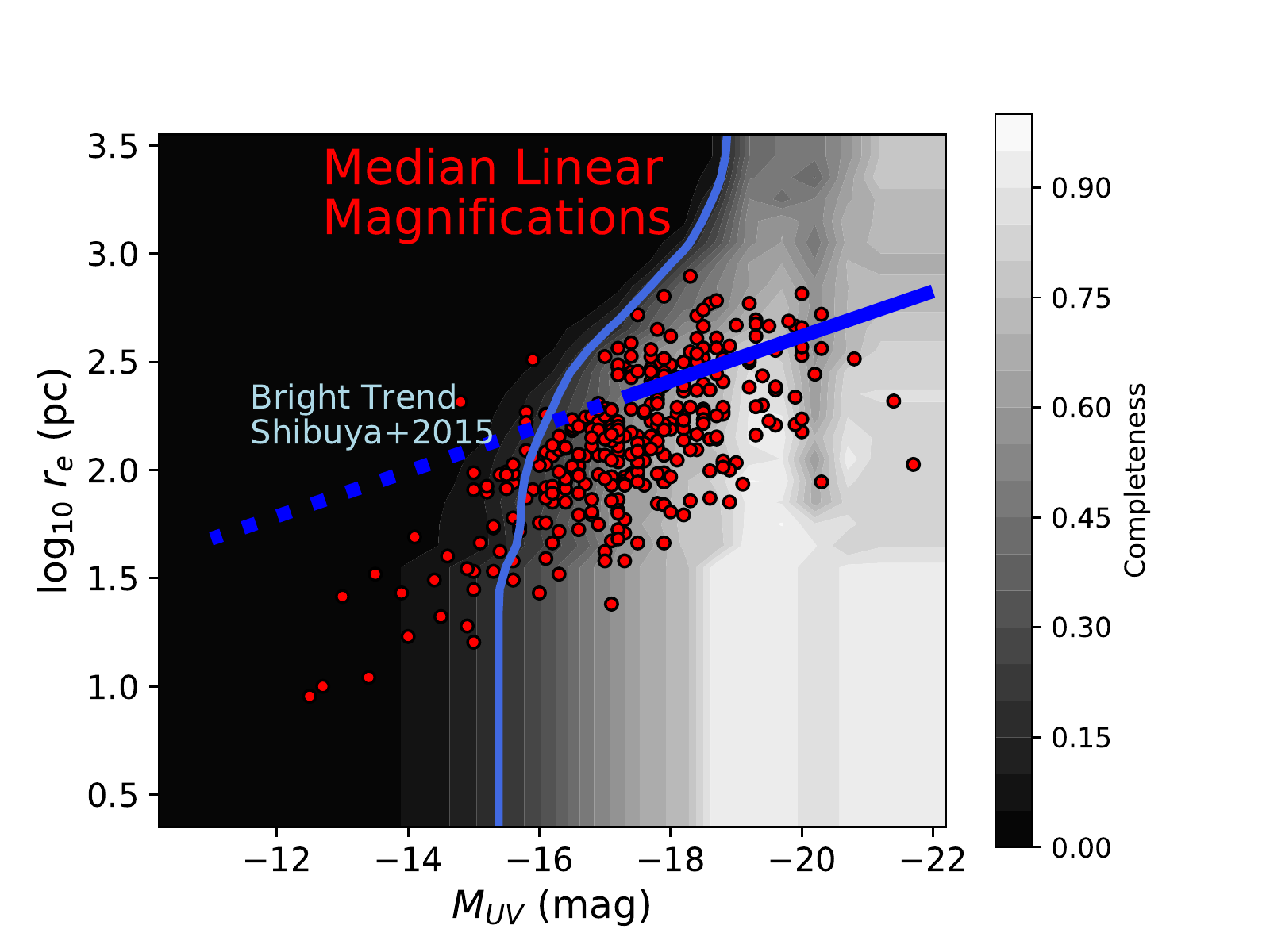}
\caption{Identifical to the left panel in Figure~\ref{fig:msre0}, but
  using a linear scaling to show the relative completeness contours.
  This shows that the $\geq$90\% of our $z\sim6$-8 sample lies at a
  relative completeness $\geq$20\% (\textit{shown with a light blue
    line}).\label{fig:sizlum_linear}}
\end{figure*}

\begin{figure*}
\epsscale{1.15}
\plotone{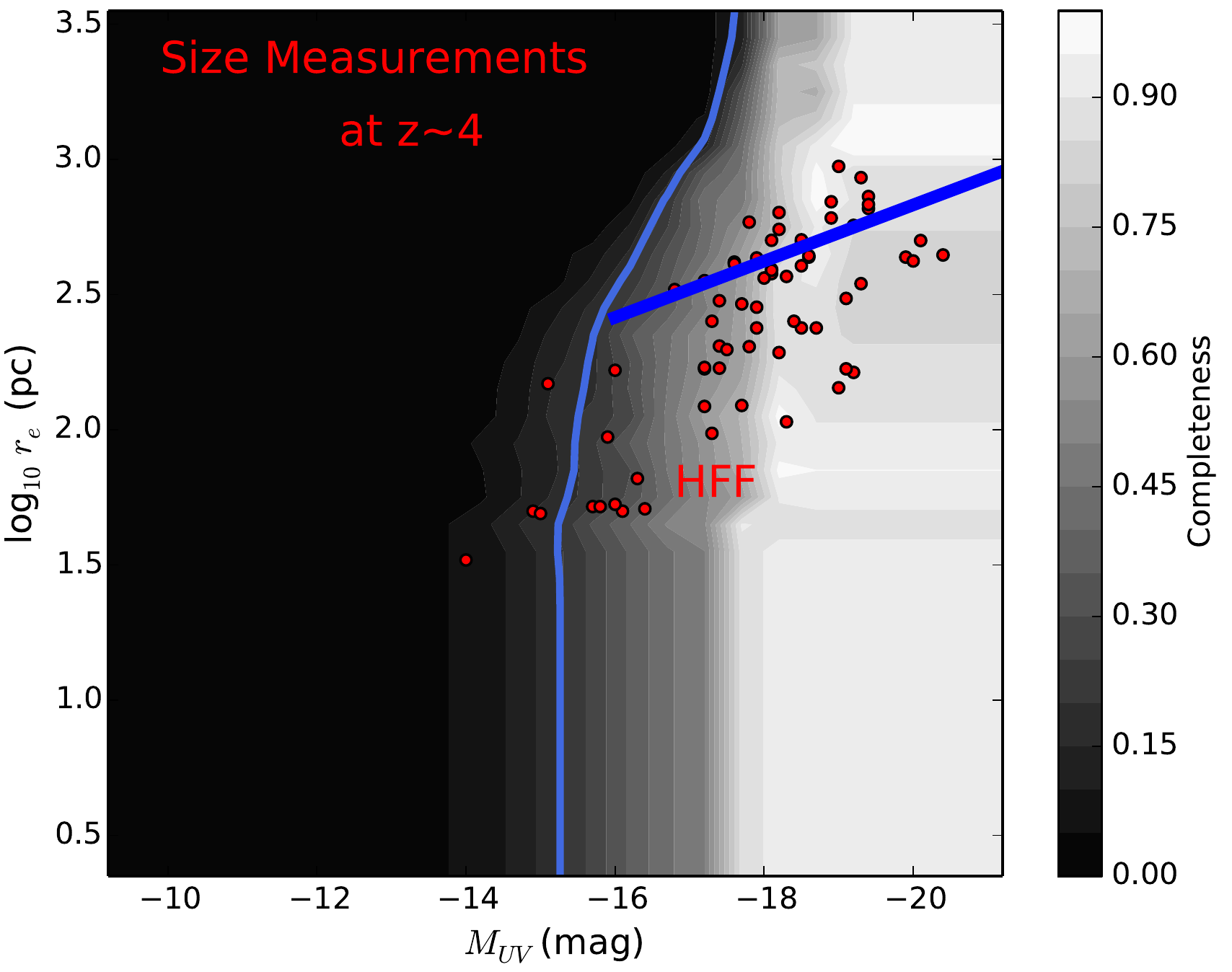}
\caption{Identifical to the left panel in Figure~\ref{fig:msre4}, but
  using a linear scaling to show the relative completeness contours.
  This shows that the $\geq$90\% of our $z\sim4$ sample lies at a
  relative completeness $\geq$20\% (\textit{shown with a light blue
    line}).\label{fig:sizlum4_linear}}
\end{figure*}

To provide us with a better sense of how sources distribute themselves
around higher completeness thresholds, i.e., $\geq 20$\%, we have
included alternative versions of the left panel to
Figure~\ref{fig:msre0} and \ref{fig:msre4} in
Figures~\ref{fig:sizlum_linear} and \ref{fig:sizlum4_linear},
respectively, using a linear scaling.  The bulk ($>$90\%) of the
sources clearly lie in regions where the relative completeness is in
excess of 20\%.

\end{document}